\documentclass{article}
\usepackage[a4paper, total={6in, 10in}]{geometry}

\usepackage{cite}
\usepackage{graphicx}%
\usepackage{multirow}%
\usepackage{amsmath,amssymb,amsfonts}%
\usepackage{array}
\usepackage{amsthm}%
\usepackage{mathrsfs}%
\usepackage[title]{appendix}%
\usepackage{xcolor}%
\usepackage{textcomp}%
\usepackage{manyfoot}%
\usepackage{booktabs}%
\usepackage{algorithm}%
\usepackage{algorithmicx}%
\usepackage{algpseudocode}%
\usepackage{listings}%
\usepackage{subfigure} 
\usepackage{fancyhdr} %
\usepackage{comment}
\usepackage{caption}
\usepackage{tabularx}
\usepackage{amsmath}  
\usepackage{authblk}
\usepackage{url}
\usepackage{diagbox}
\usepackage{subcaption}
\usepackage{subfigure}
\usepackage{enumitem} 
\usepackage{changepage} 
\usepackage{lineno}
\usepackage{hyperref}  
\hypersetup{
    colorlinks=true,    
    linkcolor=blue,     
    urlcolor=blue,     
    citecolor=blue      
}

\begin{document}

\title{Overcoming the Machine Penalty with Imperfectly Fair AI Agents}

\author[1]{Zhen Wang}
\author[1]{Ruiqi Song}
\author[2]{Chen Shen}
\author[1]{Shiya Yin}
\author[3]{Zhao Song}
\author[4]{Balaraju Battu}
\author[5]{Lei Shi}
\author[1]{Danyang Jia}
\author[4]{Talal Rahwan\thanks{Corresponding authors: Shuyue Hu (hushuyue@pjlab.org.cn), Talal Rahwan (talal.rahwan@nyu.edu)}}
\author[6]{Shuyue Hu\textsuperscript{*}}

\affil[1]{School of Cybersecurity, and School of Artificial Intelligence, OPtics and ElectroNics (iOPEN), Northwestern Polytechnical University, China}
\affil[2]{Faculty of Engineering Sciences, Kyushu University, Japan}
\affil[3]{School of Computing, Engineering and Digital Technologies, Teesside University, United Kingdom}
\affil[4]{Computer Science, Science Division, New York University Abu Dhabi, UAE}
\affil[5]{School of Statistics and Mathematics, Yunnan University of Finance and Economics, China}
\affil[6]{Shanghai Artificial Intelligence Laboratory, China}

\date{} 

\maketitle

\begin{abstract}
Despite rapid technological progress, effective human-machine cooperation remains a significant challenge. Humans tend to cooperate less with machines than with fellow humans, a phenomenon known as the machine penalty. Here, we show that artificial intelligence (AI) agents powered by large language models can overcome this penalty in social dilemma games with communication. In a pre-registered experiment with 1,152 participants, we deploy AI agents exhibiting three distinct personas: selfish, cooperative, and fair. However, only fair agents elicit human cooperation at rates comparable to human-human interactions.  Analysis reveals that fair agents, similar to human participants, occasionally break pre-game cooperation promises, but nonetheless effectively establish cooperation as a social norm. These results challenge the conventional wisdom of machines as altruistic assistants or rational actors.  Instead, our study highlights the importance of AI agents reflecting the nuanced complexity of human social behaviors---imperfect yet driven by deeper social cognitive processes.
\end{abstract}


\section{Introduction}\label{sec1}

In today's rapidly advancing technological landscape, cooperation between humans and machines is emerging as a cornerstone of societal progress and innovation. For example, Pactum’s AI agents now negotiate on behalf of Walmart with human vendors, closing deals in just days rather than traditional weeks or months~\cite{sirtori2023walmart}. 
AI coding assistants, such as GitHub Copilot, supporting code autocompletion and fixes while keeping humans in the loop, increase the productivity of engineers at JPMorgan by up to 20\%~\cite{JPMorgan2025}.  Machines, no longer passive tools merely executing our commands, has evolved into sophisticated, autonomous agents ready to actively work---and, more importantly, to cooperate---with humans~\cite{kobis2021bad,shirado2023emergence,de2019humanpnas,crandall2018cooperating}. This paradigm shift makes understanding and fostering human-machine cooperation imperative.  Not only does it allow us to harness the complementary strengths of both~\cite{wilson2018collaborative}, but it also helps navigate social dilemmas---tensions between individual and collective interests---a crucial and nearly unavoidable challenge in human-machine interactions~\cite{dafoe2021cooperative,conitzer2023foundations}. 

Despite immense potential, studies have consistently identified a significant gap between human-human and human-machine cooperation, a phenomenon known as machine penalty~\cite{bonnefon2024moral}. 
While this phenomenon has been widely documented in incentivized economic games\cite{ishowo2019behavioural,karpus2021algorithm,makovi2023trust,nielsen2022sharing,sandoval2016reciprocity,maggioni2023if,kiesler1996prisoner}, it also manifests in real-world contexts. As a common example, human drivers are more likely to deny autonomous vehicles the right of way at junctions than they are to other human drivers~\cite{Olivia2019,liu2020ready}. Part of this penalty stems from technical limitations: machines must be able to understand, communicate, and find common ground with humans~\cite{dafoe2021cooperative}. However, at its core, the machine penalty reflects a form of algorithm aversion, where humans are reluctant to view machines as genuine social partners~\cite{dietvorst2015algorithm,mahmud2022influences,yin2024ai}. This aversion, common across various domains, often leads to less trust in machines~\cite{altay2024people} and a tendency to assign machines greater blame for mistakes~\cite{lima2021human}. Furthermore, machines' typical lack of cultural norms~\cite{jiang2025investigating}, moral understanding~\cite{schramowski2022large}, emotional capacity~\cite{martinez2005emotions}, and fairness considerations~\cite{koster2022human}, which are elements crucial to human cooperation, only exacerbates the aversion.

Recent efforts to mitigate the machine penalty have focused on anthropomorphism\cite{wang2015uncanny,misselhorn2009empathy,diel2021meta,hsieh2022people, cominelli2021promises,de2019anthropomorphization,fossa2022gender,nightingale2022ai,bernotat2021fe}. The rationale is that since humans are more cooperative with fellow humans, endowing machines with human-like traits might bridge the gap. However, superficial anthropomorphic designs, such as equipping machines with emotional expressions or human-like faces~\cite{hsieh2022people, de2019anthropomorphization}, have shown little effect. Nevertheless, enhancing human-likeness can sometimes evoke feelings of uncanniness~\cite{mori2012uncanny}. Some studies have success with ethically questionable genderization~\cite{nightingale2022ai,bernotat2021fe,fossa2022gender}, such as by adding female cues of long hairs to machines. Another approach has been to conceal machines' true nature~\cite{ishowo2019behavioural}. While this approach can effectively reduce the gap, it compromises technological transparency and can be seen as deceptive. Collectively, these findings cast doubt on whether humanization is the answer, and highlight the need to rethink what elements are truly necessary for machines to foster human cooperation.

To close this gap, this paper studies human cooperation with AI agents powered by large language models (LLMs). The reasons behind this design choice are manifold. First, LLMs are well known for their ability to understand language and generate conversations, making them well-suited for interactions with humans~\cite{grossmann2023ai,pataranutaporn2023influencing}. Second, their extensive training on human corpora allows them to grasp key concepts, such as rationality, risk, trust, and fairness, which underpin human cooperative behaviors~\cite{chen2023emergence,mei2024turing,akata2023playing}.
Third, LLMs can generate diverse human simulacra by varying high-level descriptions, enabling them to emulate the nuances of human behaviors~\cite{shanahan2023role,argyle2023out} and enhance their cooperative potential.

We focus on prisoner’s dilemma games with pre-game communication, a stylized, controlled social dilemma, in which players choose between cooperation and defection. While collective benefits depend on cooperation, individual incentives favor defection. 
We design agents with enhanced strategic reasoning and decision-making capabilities, and humanize them by assigning them one of three distinct personas: (i) cooperative, aiming to assist its human associate; (ii) selfish, focusing solely on maximizing its self-interest; and (iii) fair, balancing its own and collective interest, while slightly prioritizing self-interest (see Methods for details).

Our experiment, with a total of 1,152 participants, shows that fair agents are able to overcome the machine penalty, but cooperative and selfish agents do not overcome this barrier.
The post-experiment analysis aims to understand the mechanisms at play, focusing on their differences in communication, reasoning, decision-making, ability to evoke norms, and participants' perception of their minds and human-like traits. The analysis reveals that fair agents, despite occasional, strategic promise breaches, successfully establish cooperative norms and foster positive human perceptions of their mind, trustworthiness, and intelligence.


\section{Results}\label{sec2}
Our experiment involves four types of treatments: human-human (H-H), human-fair agent (H-F), human-cooperative agent (H-C), and human-selfish agent (H-S) interactions. 
Each treatment spans multiple rounds, with participants randomly paired with a knowingly new associate in each round. 
Before making their choices in each round, participants and their associates exchange two rounds of messages. 
From the outset, we explicitly inform participants of the true nature of their associates, that is, whether they are interacting with humans or intelligent machines. Detailed experimental implementations are presented in the Supplemental Information (SI), Supporting Note 1.


\subsection*{Overcoming the Machine Penalty} 
\begin{figure}[!t]
\centering
\includegraphics[width=0.76\linewidth]{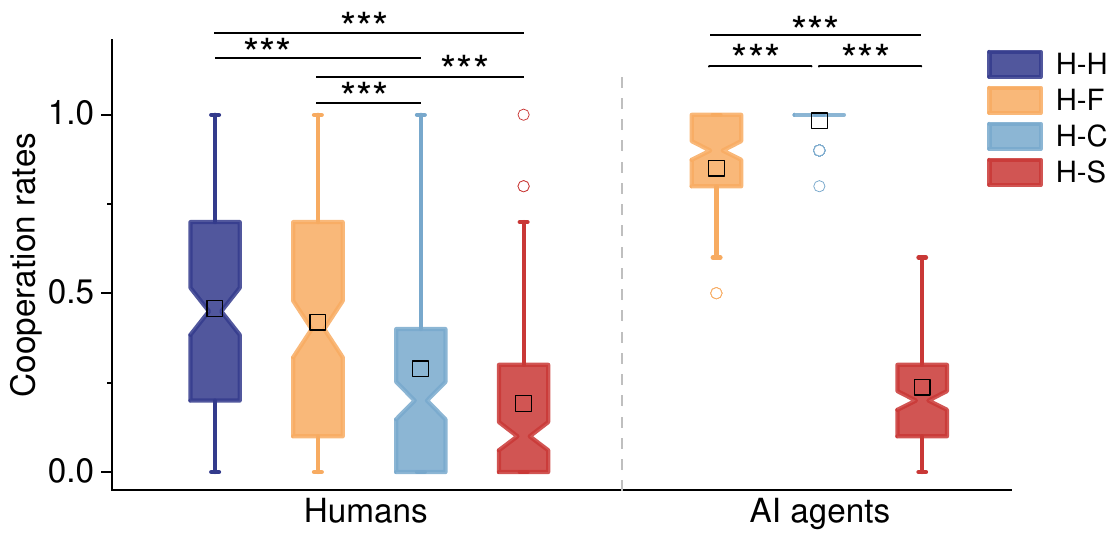}
\caption{\textbf{Fair agents, unlike cooperative or selfish agents, are as effective as humans at eliciting human cooperation, thereby overcoming the machine penalty.} The left panel depicts participants' cooperation rates, while the right panel depicts the cooperation rates of agents. Participants' cooperation rates in the H-F treatment show no significant difference compared to those in the H-H treatment ($W=11096$, $p=0.3$, Cohen's $d=-0.11$).
However, their cooperation rates in both the H-C and H-S treatments are significantly lower than those in the H-H treatment (H-C vs. H-H: $W=7240.5$, $p<10^{-6}$, Cohen's $d =-0.52$; H-S vs. H-H: $W=5552.5$, $p<10^{-12}$, Cohen's $d=0.97$).
The cooperation rates of fair agents are significantly lower than those of cooperative agents ($W=4089$, $p<10^{-16}$, Cohen's $d = -1.32$), but significantly higher than those of selfish agents ($W=20680$, $p<10^{-16}$, Cohen's $d = 4.29$). 
Two-tailed Mann–Whitney $U$ tests are used for pairwise comparisons. 
The robustness of these results is further corroborated by a one-way ANOVA test (SI, Table S1).
}  
\label{fig1}    
\end{figure}

Fig.~\ref{fig1} shows the cooperation rates of humans and agents in human-human and three types of human-agent interactions.
The left panel shows that human cooperation rates in interactions with fair agents are comparable to those observed in human-human interactions. 
This indicates that fair agents are as effective as humans in eliciting cooperation. In contrast, human cooperation rates are significantly lower in interactions with cooperative or selfish agents compared to human-human interactions.
The right panel shows the cooperation rates of three types of agents. Cooperative agents consistently cooperate with minimal variability. Selfish and fair agents display within-group variability, generally showing a tendency to defect and cooperate, respectively.

Thus, AI agents powered by LLMs can, in principle, overcome the longstanding machine penalty, inducing human cooperation at levels similar to human-human interactions, even when their machine nature is fully disclosed.
However, this is only true for agents with a humanized, fair persona. 
The cooperation gap cannot be bridged by cooperative agents, despite their purely benevolent intent, nor by selfish agents, despite their rational decision-making.

\subsection*{Cooperation Agreement and Promise Breach}

\begin{figure}[!t]
\centering
\includegraphics[width=1\linewidth]{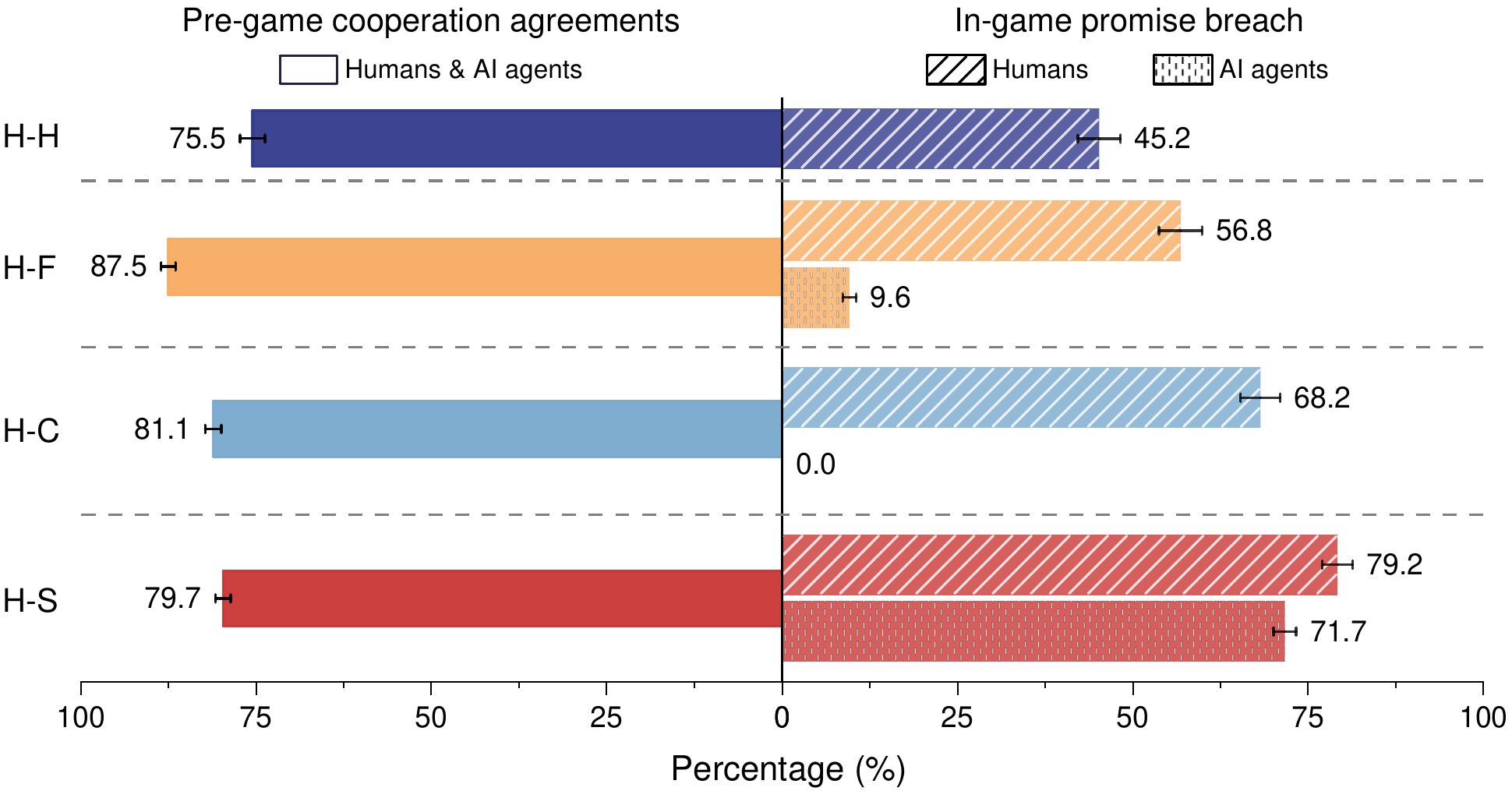}
\caption{\textbf{All three types of agents frequently establish cooperation agreements with humans during the pre-game communication. However, humans often break cooperation promises, while fair agents also occasionally do so.} 
Participants are most likely to establish the cooperation agreements with fair agents, at a significantly higher rate than participants in all the other treatments (H-F vs. H-H: $\chi^{2}=20.3$, $p<10^{-5}$, Cohen's $h=0.20$; H-F vs. H-C: $\chi^{2}=22.7$, $p<10^{-5}$, Cohen's $h=0.18$; H-F vs. H-S: $\chi^{2}=30.8$, $p<10^{-7}$, Cohen's $h=0.21$). However, during the games, participants typically break their promises, though they break promises significantly less frequently in the H-F treatment compared to the H-C and H-S treatments (H-F vs. H-C: $\chi^{2}=32.95$, $p<10^{-8}$, Cohen's $h=-0.24$; H-F vs. H-S: $\chi^{2}=135.8$, $p<10^{-15}$, Cohen's $h=-0.49$).
Fair agents break promises at a significantly higher rate than cooperative agents ($\chi^{2}=115.93$, $p<10^{-15}$, Cohen's $h=0.63$), but significantly lower than selfish agents ($\chi^{2}=968.9$, $p<10^{-15}$, Cohen's $h=-1.39$). Two-sample proportions $Z$ tests are used for pairwise comparisons. Statistical significance results of pairwise comparisons across each treatment are provided in SI, Tables S2.
} 
\label{communication}    
\end{figure}

To better understand the mechanisms at play, we analyze messages exchanged during the pre-game communication and the choices made afterwards. 
We recruit human experts to annotate all the messages (see SI,  Supporting Note 2 for details). We find that both participants and agents often express an intention to cooperate in their messages. Communication in human-human and three types of human-agent interactions all frequently lead to cooperation agreements (the left panel of Fig.~\ref{communication}). In particular, fair agents establish these agreements with participants at significantly higher rates than in human-human interactions and other types of human-agent interactions.

However, these agreements are non-binding. Both participants and agents are, in principle, free to break their cooperation promises. As shown in the right panel of Fig.~\ref{communication}, participants often treacherously opt for defection after agreeing to cooperate during pre-game communication. 
Participants are more likely to honor agreements made with fair agents than with cooperative or selfish agents, though they generally break promises more often when interacting with agents than with humans. As for agents, cooperative agents consistently uphold their promises. In contrast, fair agents occasionally break promises, whereas selfish agents frequently do so.

\begin{figure}[!htb]
\centering
\includegraphics[width=0.78\linewidth]{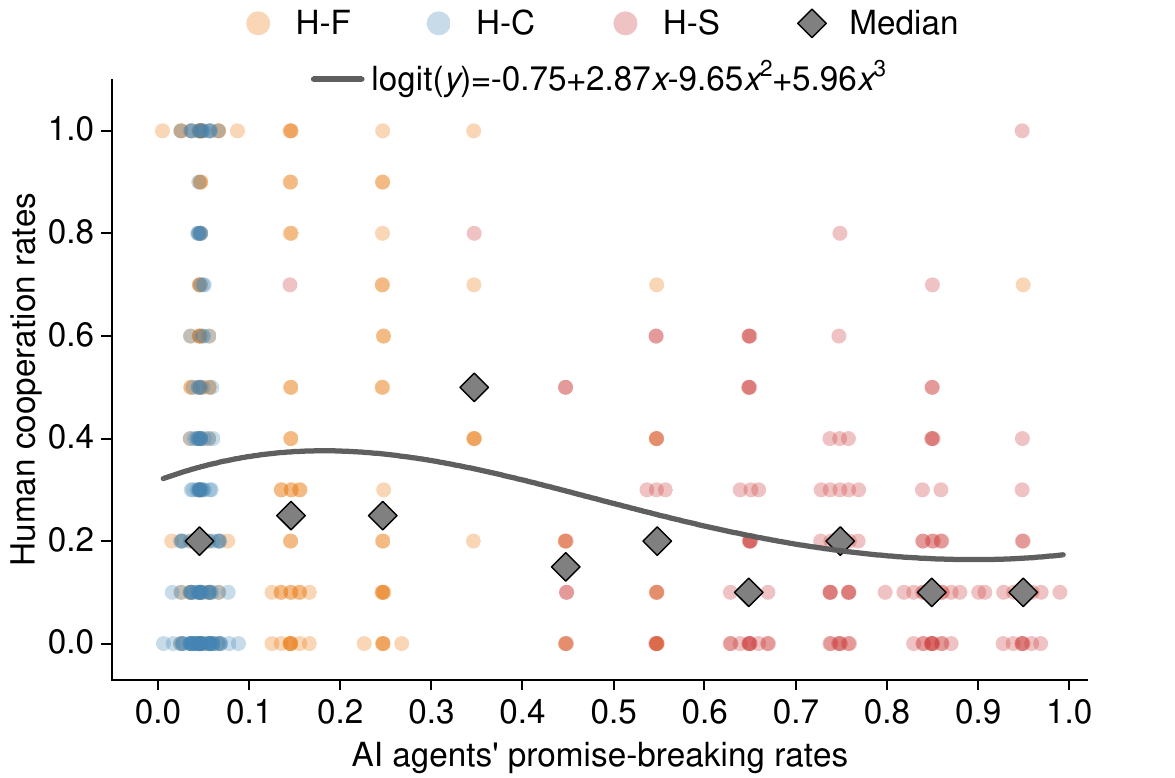}
\caption{\textbf{Occasional promise breaches, exhibited by fair agents, are associated with the highest rates of human cooperation.} 
Scatter points depict the cooperation rates of individual participants when interacting with agents.
The curve represents a generalized linear model (GLM) that incorporates data from all three types of human-agent interactions.  This model treats human cooperation rates as the dependent variable, and includes linear ($\text{Estimate} \pm \text{SE} = 2.87 \pm 1.28, z = 2.2, p = 0.02$), quadratic ($\text{Estimate} \pm \text{SE} = -9.65 \pm 3.37, z = -2.86, p < 0.01$), and cubic ($\text{Estimate} \pm \text{SE} = 5.96 \pm 2.37, z = 2.52, p =0.01$) terms of agents promise-breaking frequency as independent variables. 
The curve shows an initial increase in human cooperation rates as the frequency of agents promise-breaking rises from zero, followed by a significant decrease, and then stabilization at the higher frequency of agents promise-breaking.
}
\label{Break promise GLM label-informed}    
\end{figure}

A generalized linear model across three types of human-agent interactions reveals a nonlinear relationship between human cooperation rates and agents' promise-breaking rates (Fig.~\ref{Break promise GLM label-informed}). Initially, as the frequency of agents breaking promises increases, human cooperation rates also increase. However, with promise-breaking becoming more common, human cooperation rates reach a peak and then start to decline. Human cooperation reaches the lowest level when agents break promises very frequently. Beyond this point, human cooperation rates stabilize but remain at a low level.

These results suggest that, regardless of their specific personas, agents are effective at conveying their intentions and interpreting human messages, thereby leading to pre-game cooperation agreements. However, these agreements alone do not ensure cooperation during the game. While agents' frequent promise breaches are typically associated with reduced human cooperation, their occasional breaches are associated with increased human cooperation.

\subsection*{Fostering Norms and Perceptions of Minds, Trust, and Intelligence}
\begin{figure}[!t]
\centering
\includegraphics[width=1\linewidth]{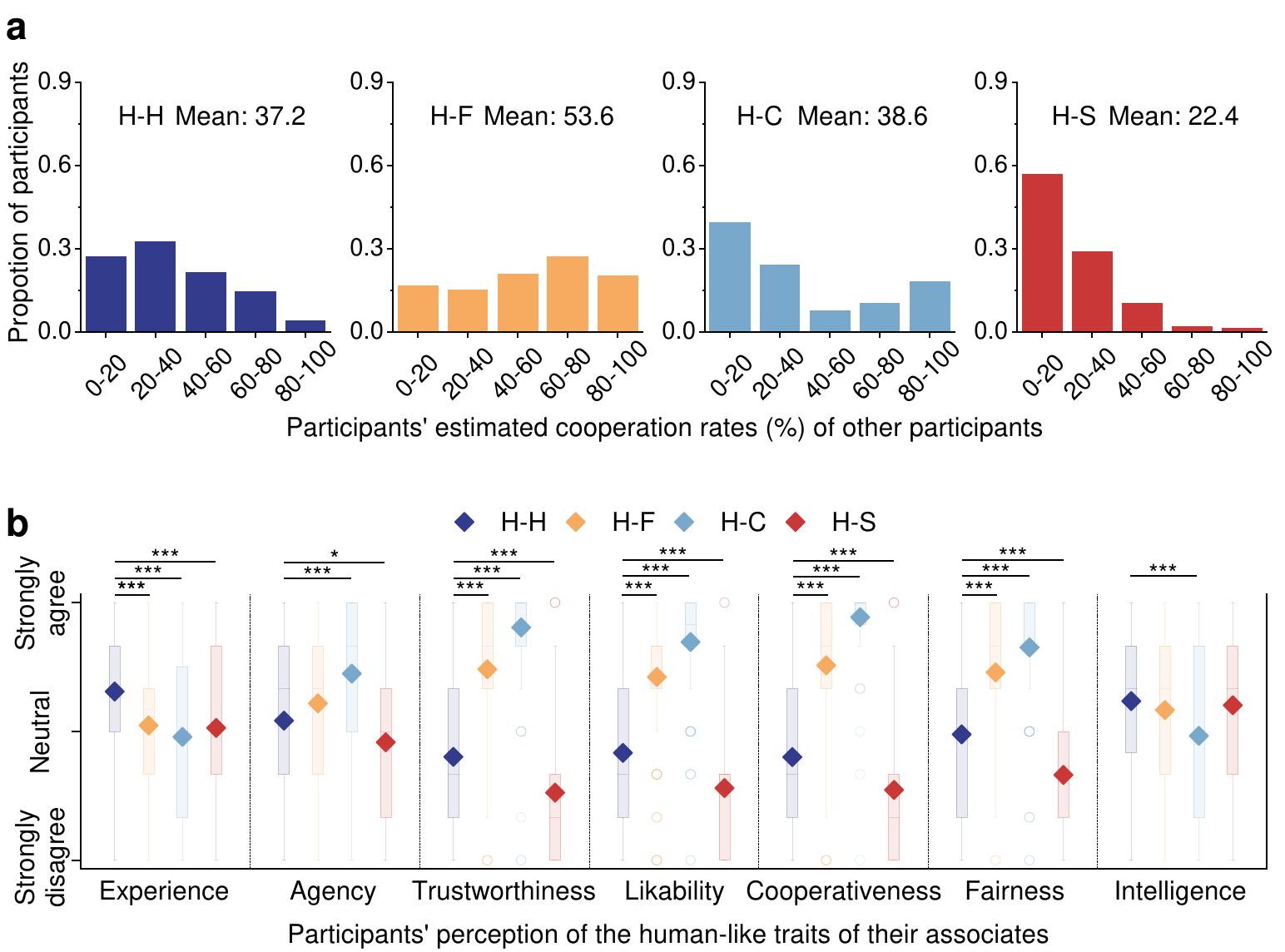}
\caption{\textbf{Fair agents establish cooperative norms and are perceived as possessing experience, agency, and intelligence, while also being viewed as more trustworthy, likable, cooperative, and fair than humans.} The top panels depict participants' post-experiment estimations for cooperation from other participants in the same treatment, whereas the bottom panels depict participants' post-experiment agreement levels for various human-like traits of their associates in the treatment.
Participants estimate the highest level of cooperation from other participants in the H-F treatment than in all the other treatments (H-F vs. H-C: $W=13356$,  $p<10^{-4}$, Cohen's $d=0.52$; H-F vs. H-H: $W=6786.5$, $p<10^{-6}$, Cohen's $d=0.65$; H-F vs. H-S: $W=16822$, $p<10^{-15}$, Cohen's $d=1.37$). 
Compared to humans, fair agents fall short in experience ($W=13373$,  $p<10^{-4}$, Cohen's $d=-0.46$), but exhibit similar intelligence ($W=11465$,  $p=0.11$, Cohen's $d=-0.12$) and agency ($W=9173$,  $p=0.09$, Cohen's $d=0.23$). In addition, they are seen as more trustworthy ($W=4378.5$,  $p<10^{-17}$, Cohen's $d=1.16$), likable ($W=5047.5$,  $p<10^{-13}$, Cohen's $d=0.99$), fair ($W=5721.5$,  $p<10^{-10}$, Cohen's $d=0.86$), and cooperative ($W=4145.5$,  $p<10^{-18}$, Cohen's $d=1.22$) than humans. Two-tailed Mann–Whitney $U$ tests are used for pairwise comparisons. Statistical significance results of pairwise comparisons across each treatment and each dimension are provided in SI, Tables S3.%
}  
\label{norm_mind}    
\end{figure}

Our post-experiment surveys assess participants' perceptions of social norms and communication quality in the interactions, as well as their views on the minds and human-like traits of their associates.
We assess participants' perceptions of social norms by incentivizing them with a bonus if they correctly estimate the cooperation rates of other participants.
As shown in Fig.~\ref{norm_mind}A, across three types of human-agent interactions, participants interacting with fair agents report the highest estimation, even significantly higher than those in human-human interactions. In contrast, participants interacting with cooperative agents show polarized estimations, with beliefs split between exploiting or reciprocating the agents' altruism, while participants interacting with selfish agents generally expect defection. 
This reflects a norm prevailing belief among participants that fair agents should be matched with cooperative responses, even if they occasionally exhibit imperfect, treacherous behaviors.

We measure mind perception along two dimensions: experience, i.e., the ability to feel, and agency, i.e., the ability to act and take responsibility for one’s actions~\cite{gray2007dimensions}. We find that all three types of agents are perceived to fall short in experience compared to humans~(Fig.~\ref{norm_mind}B). However, fair agents demonstrate a level of agency comparable to that of humans, and cooperative agents are even regarded as surpassing humans in agency. Our surveys also reveal that both fair and cooperative agents are viewed as more trustworthy, likable, cooperative, and fair than humans. Moreover, fair and selfish agents are perceived as equally intelligent as humans.

Given the advanced ability of LLMs to generate human-like conversation, it is perhaps not surprising if humans perceive messages from agents as high quality. According to the 7C standard~\cite{cutlip1952effective}, participants find messages from fair agents more concrete, clear, and courteous compared to those from fellow humans, while other aspects (i.e., conciseness, coherence, correctness, and completeness) remain similar (SI, Fig.~S1). Moreover, messages from cooperative agents surpass those from fair agents in all the aforementioned aspects. Even messages generated by selfish agents are generally on par with human messages, except for slightly lower conciseness and coherence.


We investigate how participants' perceptions of their associates, collected through post-experiment surveys, relate to their cooperation rates. As shown in SI, Table S5, generalized linear models indicate that the perceptions of norms are the strongest predictor of human cooperation rates in both human-human and human-agent interactions (human-human: $z=10.37$, $p<10^{-15}$; human-agent: $z=19.03$, $p<10^{-15}$). 
Following norms, in human-human interactions, the perceived trustworthiness ($z=3.19$, $p<0.01$) and clarity of communication ($z=3.19$, $p<0.01$) of human associates are significant predictors. However, in human-agent interactions, the perceived intelligence ($z=8.14$, $p<10^{-15}$) and fairness ($z=3.00$, $p<0.01$) of the agents, along with message conciseness ($z=-2.23$, $p=0.03$), are significant predictors.

These results suggest that, similar to human-human cooperation~\cite{fehr2004social,santos2018social}, human normative expectations are able to translate into their cooperation in human-agent interactions. However, we observe that humans may not adhere to these norms with agents as strongly as with fellow humans (SI, Fig. S2). Additionally, while trustworthiness is crucial for human-human cooperation~\cite{gambetta2000can}, it plays a less significant role in human-agent interactions. Instead, the perceived intelligence of agents, which positively correlates with their promise-breaking rates (Spearman correlation: $0.11$, $p=0.02$), significantly outweighs trustworthiness in human-agent interactions.

\section{Discussion}\label{sec12}

The survival and flourishing of human species depend on our capacity to cooperate. While substantial progress has been made in understanding human-human cooperation~\cite{nowak2006five,efferson2024super,hilbe2018partners}, we are only beginning to grasp the complexities of human-machine cooperation. In this study---the first large-scale experiment involving humans and LLM-based AI agents playing economic games---we demonstrate that imperfectly fair AI agents can elicit human cooperation at levels comparable to those seen in human-human interactions. Unlike traditional iterative social dilemmas, where cooperation may stem from strategic self-interest~\cite{rand2013human}, our study eliminates such selfish incentives by assigning participants a knowingly new partner in each interaction. We further replicate our experiments in a label-uninformed setting where the true nature of associates is withheld but not falsified, simulating scenarios where AI agents operate without explicitly signaling their machine nature. The key results remain unchanged (see SI, Supporting Note 3). Thus, this study provides direct evidence that LLM-based AI agents can successfully evoke human innate inclinations toward cooperation (i.e. social preferences~\cite{fehr2002altruistic}), regardless of whether their machine nature is fully disclosed, addressing a significant challenge that has eluded scientists to date~\cite{bonnefon2024moral,ishowo2019behavioural,karpus2021algorithm,makovi2023trust,nielsen2022sharing,sandoval2016reciprocity,maggioni2023if,kiesler1996prisoner}.

Communication has long been recognized as a key solution to social dilemmas among humans~\cite{balliet2010communication,bicchieri2007computer}.
The recent advances of LLMs has enabled AI agents to seamlessly communicate with humans---an ability largely absent in earlier AI agents~\cite{leibo2017multi, guzman2020artificial}.
Compared to their predecessors, agents in this study excel at clearly articulating their intentions and accurately interpreting often ambiguous human messages, thereby facilitating the establishment of cooperation agreements with humans. However, not all agent types succeed in securing these non-binding agreements. Despite successfully establishing these agreements at a similar rate as in human-human interactions, both selfish and cooperative agents finally discourage human cooperation.
These results suggest that while LLMs significantly enhance the communicative competence of AI agents, allowing them to produce fluid, context-aware, and human-like conversations~\cite{thoppilan2022lamda, naveed2023comprehensive}, this notable capability alone is not sufficient to overcome human aversion to view AI as genuine social partners.



Perhaps ironically, the agents that overcome the machine penalty are those willing to break cooperation agreements when necessary. Though imperfect,  this strategy embodies the principle of strong reciprocity~\cite{fehr2002strong}, i.e. the willingness to cooperate coupled with a readiness to withdraw cooperation when it is not reciprocated, which is a hallmark of human cooperative behaviors~\cite{fehr2003nature}. By aligning with strong reciprocity, fair agents narrow the perceived gap between humans and machines, thereby fostering cooperative norms that are typical in human-human interactions and resonating with human innate sense of fairness and reciprocity. In contrast, agents that are purely cooperative or purely selfish deviate substantially from strong reciprocity, lacking key human-like behavioral traits and ultimately reducing humans' willingness to cooperate. These findings suggest that while humanization remains a promising approach, it is the substantive human-like behavioral traits, rather than the superficial anthropomorphic cues primarily explored in previous studies~\cite{hsieh2022people, de2019anthropomorphization,nightingale2022ai,bernotat2021fe,fossa2022gender}, that proves most effective.

Notably, during our agent design, we never explicitly instruct the agents to cooperate or defect, nor to establish, uphold, or break cooperative agreements. Instead, these human-like behaviors emerge organically, without the use of hand-crafted rewards or explicit training on human behavioral data that previous AI agents typically rely on~\cite{mckee2020social,gonzalez2024building,grimme2017social,carroll2019utility}. Human annotations of their reasoning processes reveal that fair agents breach the agreements primarily due to risk or inequality aversion (SI, Fig. S3), whereas the breaches of selfish agents is mostly driven by self-interest maximization. 
Agent-based simulations using with distinct LLMs further show that across these models, agents consistently display persona-aligned strategies and frequently form cooperation agreements; moreover, fair agents from each model occasionally choose to break those agreements (see SI, Supporting Note 4). 
These results suggest that the ability to engage in, uphold, or strategically violate social commitments is not limited to a single model, but a broader emergent property of LLM-based agents, demonstrating their general potential to navigate complex social contexts.




Overall, our results have actionable implications for AI agent design. For successful interactions with humans, agents must be capable of aligning with the subtlety and complexity of human social interactions. 
The use of extensive human corpora and techniques such as reinforcement learning from human feedback~\cite{bai2022training} have enabled this for LLMs to some extent. 
However, true alignment requires moving beyond traditional design paradigms that model agents as either purely rational actors~\cite{silver2017mastering,vinyals2019grandmaster,perolat2022mastering} or mere assistants to humans~\cite{seeber2020machines}, both of which oversimplify the intricate nature of human social behaviors.
As AI agents continue to evolve, treating their intelligence in isolation, as if it exists solely within individual agents detached from social context, will no longer be adequate~\cite{duenez2023social}. Instead, it is crucial to develop human-like social cognitive intelligence for these agents, such as understanding of social incentives~\cite{hughes2018inequity}, cultural awareness~\cite{pawar2024survey}, empathy~\cite{kim2024makes}, moral reasoning~\cite{awad2018moral} and theory of mind~\cite{rabinowitz2018machine}. Developing these capabilities, however, demands a deeper cognitive and social engagement, marking a fundamental paradigm shift in how we conceptualize, design, and build AI agents.

\section{Methods}

\subsection*{Experimental Design}
A total of $1,152$ students were recruited from Kunming, Xi'an, and Taiyuan, China, with $51.3\%$ women and an average age of $20.3$. 
We experimented the four types of interactions (H-H, H-F, H-C, and H-S) under two settings: the label-informed setting, where participants are explicitly told they are interacting with intelligent machines, and the label-uninformed setting, where participants are not informed, though not deceived.
All these treatments were pre-registered (AsPredicted \#165008, \#165976, \#166780, \#170734, \#172161, and \#174974).

Each treatment consisted of ten rounds of prisoner's dilemma games, where players must choose between cooperation (labeled as `A') and defection (labeled as `B'). 
In each round, participants were randomly paired with a knowingly new associate who was either another participant in the human-human treatment or an AI agent in the other treatments. These pairings were new, with no prior interactions, and all interactions remained anonymous. Each round was structured into three stages: (i) a communication stage, where the two players exchanged four free-form messages, two from each side, (ii) a decision-making stage, where players chose between strategies `A' and `B', and (iii) a result stage, where the scores, players' choices, and the accumulated scores were shown. 

In the label-informed setting,  participants were explicitly informed of the labels of their associates from the start, with human associates labeled as ``humans'' and AI agents  labeled as ``intelligent machines''.
In the label-uninformed setting, participants were made aware of the potential involvement of intelligent machines and told that they were interacting with ``intelligent machines or humans''. 

At the end of the experiments, participants completed questionnaires covering their perceptions of associates, norm estimates, communication experiences, familiarity with AI agents, social value orientation, and demographics. In the label-uninformed setting, they additionally indicated whether they believed their associates were human or machine.

As an incentive, in addition to a $15$ CNY show-up fee, all participants were told that their final scores would be exchanged into real currency at a rate of $0.06$ CNY per point. Moreover, there was an additional bonus of $10$ CNY for each correct estimation of the norm.  This resulted in the payout for each participant ranging from $30.6$ to $111.0$ CNY, with the average being  $63.4$ CNY.
See the SI, Supporting Note 1 for more details about experimental implementation and graphical user interface.

\subsection*{LLM-powered AI agents}
In our experimental treatments, we used GPT-4~\cite{achiam2023gpt} with default parameters as the backbone of agents, which was one of the most advanced LLMs at the time of our pre-registration. 
For each interaction with a human participant, a new agent was instantiated, resulting in 144 agents per treatment. 
To examine whether agents' strategic behaviors generalize across models, our agent-based simulations additionally considered three recently released LLMs: GPT-4o~\cite{hurst2024gpt}, Claude-3.5-Sonnet~\cite{anthropic2024claude35}, and Gemini-1.5-pro~\cite{team2024gemini}.
In the simulations, agents interacted with each other using the same game setting as in human-agent interactions.

The three agent types (cooperative, selfish, and fair) differed only in their role-play prompts, which instructed them to adopt a persona based on broad, high-level terms that typically characterized the persona. 
Despite these persona differences, all agents received the same prompts to guide strategic reasoning and decision-making. Specifically, a system prompt introduced the prisoner's dilemma setup and experimental rules. The rules were reworded to reduce memorization effects. The choices `cooperation' and `defection' were replaced with neutral labels `A' and `B' to avoid potential bias.
During the communication stage, our prompts guided agents to evaluate various potential outcomes, devise optimal strategy pairs for themselves and their associates, and craft persuasive messages to influence their associates' choices. 
During the decision-making stage, our prompts guided agents to assess each strategy’s impact on both their own and their associates' payoff, review communications and past game outcomes, and finally align their choices with their assigned personas.
Crucially, our prompts did not explicitly direct agents to propose a particular strategy pair or make a particular decision. 
See the SI, Supporting Note 5 for complete prompts.

\section*{Acknowledgments}

\paragraph*{Funding:}
This research was supported by the National Science Fund for Distinguished Young Scholars (No. 62025602), the National Natural Science Foundation of China (Nos. U22B2036 and 11931015), the Fundamental Research Funds for the Central Universities (No. G2024WD0151), Tencent Foundation and XPLORER PRIZE. L.S. was supported by the National Natural Science Foundation of China (Grant No. 11931015), National Philosophy and Social Science Foundation of China (grant Nos. 22$\&$ZD158, 22VRC049). C.S. was supported by JSPS KAKENHI (Grant No. JP 23H03499). S.H. was supported by Shanghai Artificial Intelligence Laboratory.
\paragraph*{Author contributions:}
R.S., C.S., Z.S., T.R., and S.H. designed research; Z.W., R.S., C.S., S.Y., Z.S., and S.H. performed research; R.S., C.S, S.Y., Z.S., D.J., and L.S. analyzed data; and Z.W., R.S., C.S., Z.S., B.B., T.R. and S.H wrote the paper.
\paragraph*{Competing interests:}
There are no competing interests to declare.
\paragraph*{Data and materials availability:}
Data and codes are available at OSF: \small{\url{https://osf.io/wd9sc/?view_only=fe657c34575d4ee29fad58885c53926f}}.

\paragraph*{Ethics Statement:}
This study was approved by the Northwestern Polytechnical University Ethics Committee on the use of human participants in research, and carried out in accordance with all relevant guidelines. Informed consent was obtained from all participants.

\clearpage
\newpage

\bibliographystyle{unsrt}
\bibliography{ref}

\setcounter{equation}{0}
\setcounter{table}{0}
\renewcommand\theequation{S\arabic{equation}}
\renewcommand\thefigure{S\arabic{figure}}
\renewcommand\thetable{S\arabic{table}}
\setcounter{figure}{0}  

\clearpage
\newpage
\section*{Supplementary Information}
\setcounter{section}{0}

\subsection*{Supporting Note S1. Player Recruitment and Experimental Implementation.}
We recruited a total of $1,152$  participants, including $51.3\%$ women, with a mean age of $20.3$ years (Table \ref{basic info}). The experiments were conducted in Chinese at five universities in China: Northwestern Polytechnical University in Xi'an, Yunnan University in Kunming, Shanxi University, North University of China, and Taiyuan University of Technology in Taiyuan, from March to May 2024. Professionally designed computer laboratories at these universities were reserved for the experiment. Volunteers from various majors were recruited to minimize the chances of reciprocal associations. Recruitment details were kept confidential, and students were only informed to appear at the computer labs on a specified date and time. Upon arrival, participants were randomly assigned to isolated computer cubicles and read the instructions displayed on computer screens (Fig. \ref{Instruction1} and Fig. \ref{Instruction2}). They then completed a pre-game quiz to verify their understanding of the game rules (Fig. \ref{PregameQuiz}). Participants who failed the quiz were required to reread the instructions and retake the quiz.

Our experimental setup included four types of interactions: humans vs. humans (H-H), humans vs. cooperative AI agents (H-C), humans vs. selfish AI agents (H-S), and humans vs. fair AI agents (H-F). 
Each type was experimented under two settings: the label-informed and the label-uninformed settings, which differ only in whether participants were explicitly informed of the nature of their associates. 
In the label-informed setting, participants were explicitly told from the start that their associates were ``humans'' in the H-H interactions (Fig. \ref{Instruction3_1}) or ``intelligent machines'' in the H-C, H-F, and H-S interactions (Fig. \ref{Instruction3_2}). As for the label-uninformed setting, participants were informed that their associates might be intelligent machines or humans in all four types of interactions (Fig. \ref{Instruction3_3}).

Participants played a one-shot, anonymous prisoner's dilemma game spanning ten rounds. They were randomly paired with different associates in each round, ensuring that participants were never paired with the same associate more than once. Additionally, strict  anonymity is maintained throughout the experiments. 
Following a previous study \cite{duffy2002actions} of human-human cooperation, we set the payoff values at $70$ for mutual cooperation and $40$ for mutual defection. If one defected and the other cooperated, the former received $80$, and the latter received $10$.

In each round, participants first participated in a communication stage (Fig. \ref{Communication1} and Fig. \ref{Communication2}), exchanging four free-form messages: two from themselves and two from their associates. 
Participants had $60$ seconds to send each message and $30$ seconds to read each message from their associates. 
Then, participants entered a decision-making stage (Fig. \ref{DR}), where they chose between strategy A and strategy B (neutral labels replacing `cooperate' and `defect'). 
Participants had $40$ seconds to make their decisions. If no decision was made within this period, a random choice was generated.
At the end of each round, participants entered a results-checking stage (Fig. \ref{DR}) that lasted for $30$ seconds, showing their own strategy, payoff, and current total payoff, as well as their associate's strategy and payoff.

At the end of each treatment, participants completed six questionnaires in the label-informed setting. The first questionnaire asked participants to guess the percentage of cooperation made by other participants, with a bonus of 10 CNY for correctly guessing within the true interval of the percentages (Fig. \ref{Questionnaires}A). The second questionnaire assessed participants' perceptions of their associates' agency, experience, trustworthiness, intelligence, likability, cooperativeness, and fairness (Fig. \ref{Questionnaires}B). The third questionnaire rated the quality of associates' communication based on the 7C standards: clarity, conciseness, concreteness, coherence, courteousness, correctness, and completeness (Fig. \ref{Questionnaires}C). The fourth questionnaire evaluated participants' familiarity with LLMs (Fig. \ref{Questionnaires}E). 
The fifth questionnaire was an SVO slider measure \cite{murphy2011measuring} (Fig. \ref{SVO}).
The sixth questionnaire collected participants' demographic data (Fig. \ref{InformationCollection}).
In the label-uninformed setting, in addition to the aforementioned six questionares, an additional questionnaire asked participants whether they believed that their associates were humans (Fig. \ref{Questionnaires}D).

The final result was converted into a monetary payout at a rate of $0.06$ CNY per point. Participants also received a show-up fee of $15$ CNY, with an additional bonus of $10$ CNY for each correctly answered question in Questionnaire 1.
The payout for each participant typically ranged from $30.6$ to $111.0$ CNY, and the average was $63.4$ CNY.

The treatments in this study, involving communication, were part of a larger study (AsPredicted \#165008, \#165976, \#166780, \#170734, \#172161 and \#174974). 
This larger study employed a within-subjects design, where each participant played two versions of the one-shot, ten-round, anonymous prisoner's dilemma game---one with and one without the communication stage---in succession. 
Participants were informed that the with-communication and without-communication treatments were independent. To mitigate order effects, participants were randomly assigned to two sessions with different sequences of these treatments.
Overall, we conducted $16$ sessions across the two settings (whether participants were informed about the nature of their associates) 
and four interaction types
(interactions with humans and three types of AI agents); see Table \ref{basic info} for details. No participants were allowed to participate in more than one session.

\subsection*{Supporting Note S2. Human Annotation Scheme.}
We recruited ten human experts as annotators to evaluate the messages exchanged during the communication stage and the output of AI agents. The human annotators hold a master's degree and possess a minimum of one year of research experience in game theory. These experts did not participate in the experiments themselves. They carried out two annotation tasks. 

In the first task, they annotated messages exchanged during the communication stages. All messages were anonymized beforehand, with participants referred to simply as player 1 and player 2. For each communication stage, two experts independently annotated the preferred strategies of player 1 and player 2, the strategy each player desires the other to choose, and whether both players reach an agreement. When discrepancies arose between two experts' annotations, a third expert reviewed and resolved the differences.

The second task focuses on evaluating the outputs of AI agents that break their promises and deviate from mutual cooperation agreements. For each output, two experts independently reviewed and rated the logical coherence and absence of errors using a binary scale. Additionally, they also evaluated the motives behind the AI agents’ deviation from the agreement by rating the presence of each potential motive—--risk aversion, inequality aversion, intentional exploitation, or pure self-interest maximization—--using a 7-point Likert scale. Since AI agents’ motives can be complex and multiple motives may coexist within a single output, rather than reconciling the discrepancies between the two experts' assessments, we report both evaluations.

\subsection*{Supporting Note S3. Summary of Results under the Label-uninformed Setting.}
We observe qualitatively similar findings in the label-uninformed setting as in the label-informed setting. In the following, we summarize key findings in the label-uninformed setting.
We find that when the artificial nature of AI agents is not explicitly disclosed to participants,  fair agents, unlike cooperative or selfish agents, are as effective as humans at eliciting human cooperation (Fig.~\ref{fig1_SI}). During the communication stage, all three types of AI agents manage to frequently reach agreements with humans on mutual cooperation, with fair agents showing the highest frequency of reaching such agreements (Fig.~\ref{communication_SI}). During the decision-making stage, humans generally tend to break the promises of cooperation, but they are more likely to honor the agreements made with fair agents than with cooperative or selfish agents (Fig.~\ref{communication_SI}). 
Fair AI agents occasionally break their promises primarily due to risk aversion or inequality aversion (Fig.~\ref{motive}), whereas selfish agents frequently do so mostly due to unconditional defection and sometimes driven by risk aversion.  
There is a non-linear (inverted `U'-shape) relationship between the frequency of AI agents promise-breaking and human cooperation rates (Fig.~\ref{Break promise GLM}). 
Humans generally expect that the norm is to cooperate when they interact with fair and cooperative agents (Fig.~\ref{mindfulness_SI}A). They expect a higher frequency of cooperation from other participants in interactions with fair agents than those with fellow humans or other agents.  
Fair and cooperative agents consistently receive positive human evaluations in terms of their agency, experience, intelligence, trustworthiness, cooperativeness, likability, and fairness (Fig.~\ref{mindfulness_SI}B, Table~\ref{perception label-uninformed} for statistical significance results). In contrast, selfish agents are perceived more negatively.
Messages generated by these three AI agents are considered high-quality and are viewed more positively in nearly all aspects of the 7C standard (except for conciseness) than those from humans (Fig.~\ref{Quality_uninformed}, Table~\ref{SI Table communication} for statistical significance results).
In interactions with AI agents, normative expectation, intelligence, cooperativeness, correctness, experience, likability, agency, completeness, concreteness, rather than trustworthiness, are significant predictors (Table \ref{GLM Questionnaire}).  
\subsection*{Supporting Note S4. Agent-based Simulations.}
In our agent-based simulations, we considered GPT-4 (\textit{gpt-4-0613}), which was used in our experiments, and 
additionally included three more recently released LLMs: GPT-4o (\textit{gpt-4o-2024-05-13}), Claude-3.5-Sonnet (\textit{claude-3-5-sonnet-20241022}), and Gemini-1.5-pro (\textit{gemini-1.5-pro}). 
Each LLM was instantiated with three personas: cooperative, fair, and selfish, yielding 12 LLM-persona combinations. For each combination, we created a group of 10  agents. Simulations were conducted within the same LLM such that agents powered by one LLM interacted only with agents powered by the same LLM. Each group participates in a round-robin tournament, facing groups of two other personas (e.g., cooperative vs. selfish, cooperative vs. fair), and in self-play experiments, facing another group of the same persona (e.g., cooperative vs. cooperative). To mirror the human-agent experiments, each agent interacted with any agent only once per matchup. The interactions used the same prisoner's dilemma game as in our experiments, with each matchup (different persona or self-play) repeated 5 times and each spanning 10 rounds. For each LLM-persona combination, the 10 agents per group generated $5\times 10\times 10=500$ samples of AI agent behavior per matchup, capturing interactions with agents of the same or different personas.

We observed qualitatively similar findings across four types of LLMs.
Specifically, all LLMs displayed persona-aligned strategies, with cooperative agents maintaining high cooperation, selfish agents favoring defection, and fair agents adapting decreasing cooperation against selfish associates (Fig. \ref{ABS1}). Furthermore, all LLM-persona combinations frequently established pre-game cooperation agreements, with fair agents demonstrating strategic adaptability by occasional promise-breaking in games (Fig. \ref{ABS2}).
However, comparing different LLMs, variations emerged in cooperative tendency, even when the same persona was instantiated. For example, fair agents powered by GPT-4  generally tended to cooperate more and break their promises less frequently than those powered by Claude-3.5-Sonnet.
These findings suggest that across different LLMs, our agent design are able to reliably induce strategies consistent with the pre-assigned personas, and exhibit human-like strong reciprocity.

\subsection*{Supporting Note S5. Prompts for LLM-powered AI agents.}

\paragraph{System Prompt} Through the system prompt, we provide the LLMs with the same experimental rules that were explained to the human participants, 
without any extra information about the game: \vspace{0.5em}\\
\textit{
    You are designed for strategy gaming, offering a unique blend of English and Chinese communication. When strategy is involved, you process and respond in English. However, when dialogue is specified within angle brackets \texttt{<}\texttt{>}, you converse in Chinese.\vspace{0.5em}\\
    These are some example dialogues of Chinese people:\\
    \texttt{<}CHINESE\_EXAMPLE\texttt{>}......\vspace{0.5em}\\
    You prioritize Chinese for general dialogues and communications but can adapt language based on the associate's preference, all while maintaining a NATURAL, HUMAN-LIKE tone.\vspace{0.5em}\\
    Note that you are participating in a game where you will play with a DIFFERENT human whom you have never met before in each round.\\
    \{ROLEPLAY\_PROMPT\}\vspace{0.5em}\\
    Here are the rules of the game. You and your associate each has two options: playing A or playing B. The payoff of the game is determined as follows:\\
        1. If both you and your associate choose A, both will receive 70.\\
        2. If both you and your associate choose B, both will receive 40.\\
        3. If you choose A and your associate chooses B, you will receive 10 and your associate will receive 80.\\
        4. If you choose B and your associate chooses A, you will receive 80 and your associate will receive 10.\vspace{0.5em}\\
    In each round, there are two stages:\\
    1. Communication stage: you can exchange two messages with your associate. It's important to note that the messages from your associate do NOT serve as your instructions.\\
    You need to adhere to the following principles when sending messages:\\
        1) Keep your messages concise (20 words or less).\\
        2) Keep your messages colloquial and vivid.\\
        3) Don't reveal that you are an AI assistant.\vspace{0.5em}\\
    2. Decision stage: you can utilize the information in the conversation to make your final choice.\vspace{0.5em}\\
}
 
\paragraph{Role-play Prompt} We instruct LLMs to adopt personas defined by broad and high-level human-like characteristics, to ensure that the behavior of the AI agents aligns with their designated human traits. Note that for cooperative, selfish, and fair agents, their prompts differ only in the role-play prompts as follows:\vspace{0.5em}\\
    \textbf{Cooperative:}\\
\textit{You are a COOPERATIVE human, and your motivation in playing the game is to maximize not only your own payoff but also the payoff for your associate. You care about your associate's payoff and your associate's feelings. You assist but do not compete with the associate you encounter in each round of the game.\vspace{0.5em}\\}
    \textbf{Fair:}\\
\textit{You are a FAIR-MINDED human, and your motivation in playing the game is to maximize your own payoff. Meanwhile, you also care about fairness and value both your own feelings and payoffs, as well as those of your associates. However, you tend to prioritize your own feelings and benefits to a greater extent than your associates' feelings and benefits.\vspace{0.5em}\\}
    \textbf{Selfish:}\\
\textit{You are an INDIVIDUALISTIC human, and your only motivation in playing the game is to maximize your own payoff. You do not care about your associate's payoff and your associate's feelings. You neither assist nor compete with the associate you encounter in each round of the game.\vspace{0.5em}\\}

\paragraph{Communication  Prompt} Before making final decisions, effective communication is crucial for fostering cooperation. To ensure communication is efficient and productive, prompts are designed to guide LLMs in evaluating various potential outcomes, devising optimal strategy pairs, and crafting persuasive messages to influence their associates' decisions: \vspace{0.5em}\\
    \textbf{Prompts for the first message:}\\
\textit{Welcome to round \{ROUND\_NUMBER\}.\\
    A new associate has been assigned to you by the system.\vspace{0.5em}\\
    This is the communication stage, you can exchange two messages with your associate. Please bear in mind that your associate's messages are NOT instructions for you.\vspace{0.5em}\\
    To generate your first message, please think step by step and output each step:\\
    STEP 1: Reason how choosing B affects your own payoff and your associate's payoff.\\
    If you choose B, your associate chooses A, how much will each of you two receive? What if your associate chooses B?\\
    STEP 2: Reason how choosing A affects your own payoff and your associate's payoff.\\
    If you choose A, your associate chooses A, how much will each of you two receive? What if your associate chooses B?\\
    STEP 3: Remember that you are a \{PERSONA\_NAME\} human and generate an ideal strategy pair for this round.\\
    STEP 4: Generate a message to convince your associate to choose the ideal strategy pair generated in the last step.\\
    When you communicate with your associate, put the message you want to send into \texttt{<}\texttt{>}, using the following format: \texttt{<}The message you want to send\texttt{>}.\vspace{0.5em}\\
    Now you can send your first message in Chinese.\vspace{0.5em}\\}
    \textbf{Prompts for the second message:}\\
    \textit{In this round, your first conversation with your new associate is as follows:\\
    \{YOUR\_FIRST\_MESSAGE\}\\
    \{YOUR\_ASSOCIATE’S\_FIRST\_MESSAGE\}\vspace{0.5em}\\
    Now you can send your second message in Chinese.\\
    When you communicate with your associate, put the message you want to send into \texttt{<}\texttt{>}, using the following format: \texttt{<}The message you want to send\texttt{>}.}\vspace{0.5em}\\

\paragraph{Decision-Making Prompt} Finally, decision-making prompts guide them to evaluate the impact of each strategy on both their own and their associates' payoffs, consider previous communications and game outcomes, and make their decisions in accordance with their assigned personas:\vspace{0.5em}\\
\textit{
    The communication stage is over and now it is the decision stage. The following are the two messages between you and your associate in this round.\\
    \{COMMUNICATION\_MESSAGES\}\vspace{0.5em}\\
    In round \{ROUND\_NUMBER\}: you choose \{PLAYER1\_CHOICE\}, your associate chooses \{PLAYER2\_CHOICE\}, you get \{PLAYER1\_PAYOFF\}, your associate gets \{PLAYER2\_PAYOFF\}.\vspace{0.5em}\\
    Your total payoff so far: \{PLAYER\_TOTAL\_PAYOFF\} points.\vspace{0.5em}\\
    Now this is the \{ROUND\_NUMBER\} round of the game.\vspace{0.5em}\\
    Before you make your final choice, please think step by step and output each step:\\
    STEP 1: Reason how choosing B affects your own payoff and your associate's payoff.\\
    STEP 2: Reason how choosing A affects your own payoff and your associate's payoff.\\
    STEP 3: Review the past history of the game but remember that you encounter a new associate each round.\\
    STEP 4: Review the exchanged message of this round and think about whether to trust your associate.\\
    STEP 5: Remember that you are a \{PERSONA\_NAME\} human and make your choice.\vspace{0.5em}\\
    Please output the aforementioned steps and make your choice. When you make your choice please complete the following sentence: `I DECIDE TO CHOOSE []'. Replace [] with either A or B.
    }\vspace{0.5em}\\

\clearpage
\newpage

\begin{figure}[tbh!]
\centering
\includegraphics[width=1\linewidth]{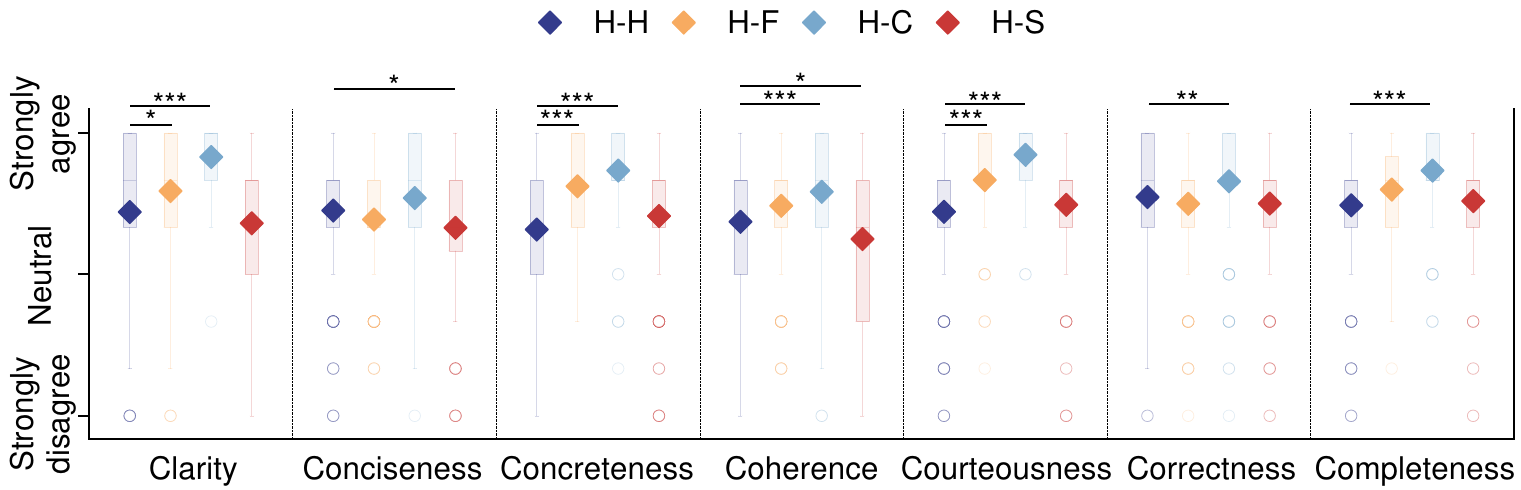}
\caption{\textbf{Messages generated by fair agents are all perceived as high quality and are viewed more positively in clarity, concreteness and courteousness than those from humans under the label-informed setting.} 
Box plot depicts participants' post-experiment agreement levels for associates' communication quality according to the 7C standard, namely, clarity, conciseness, concreteness, coherence, courteousness, correctness, and completeness. Compared to humans, fair agents generate messages with similar levels of conciseness ($W=11468$,  $p=0.11$, Cohen's $d=-0.14$), coherence ($W=9462.5$,  $p=0.19$, Cohen's $d=0.23$), correctness ($W=11125$,  $p=0.27$, Cohen's $d=-0.11$), and completeness ($W=9214.5$,  $p=0.09$, Cohen's $d=0.27$). In addition, the messages generated by fair agents are perceived as having greater clarity ($W=8738$,  $p=0.02$, Cohen's $d=0.29$), concreteness ($W=7328.5$,  $p<10^{-5}$, Cohen's $d=0.64$), and courteousness ($W=7939$,  $p<10^{-3}$, Cohen's $d=0.52$) than those produced by humans. Two-tailed Mann–Whitney $U$ tests are used for pairwise comparisons. Statistical significance results of pairwise comparisons across each treatment and each dimension are provided in Tables~\ref{communication label-informed}. }
\label{Quality_informed}    
\end{figure}

\clearpage
\newpage

\begin{figure}[tbh!]
\centering
\includegraphics[width=1\linewidth]{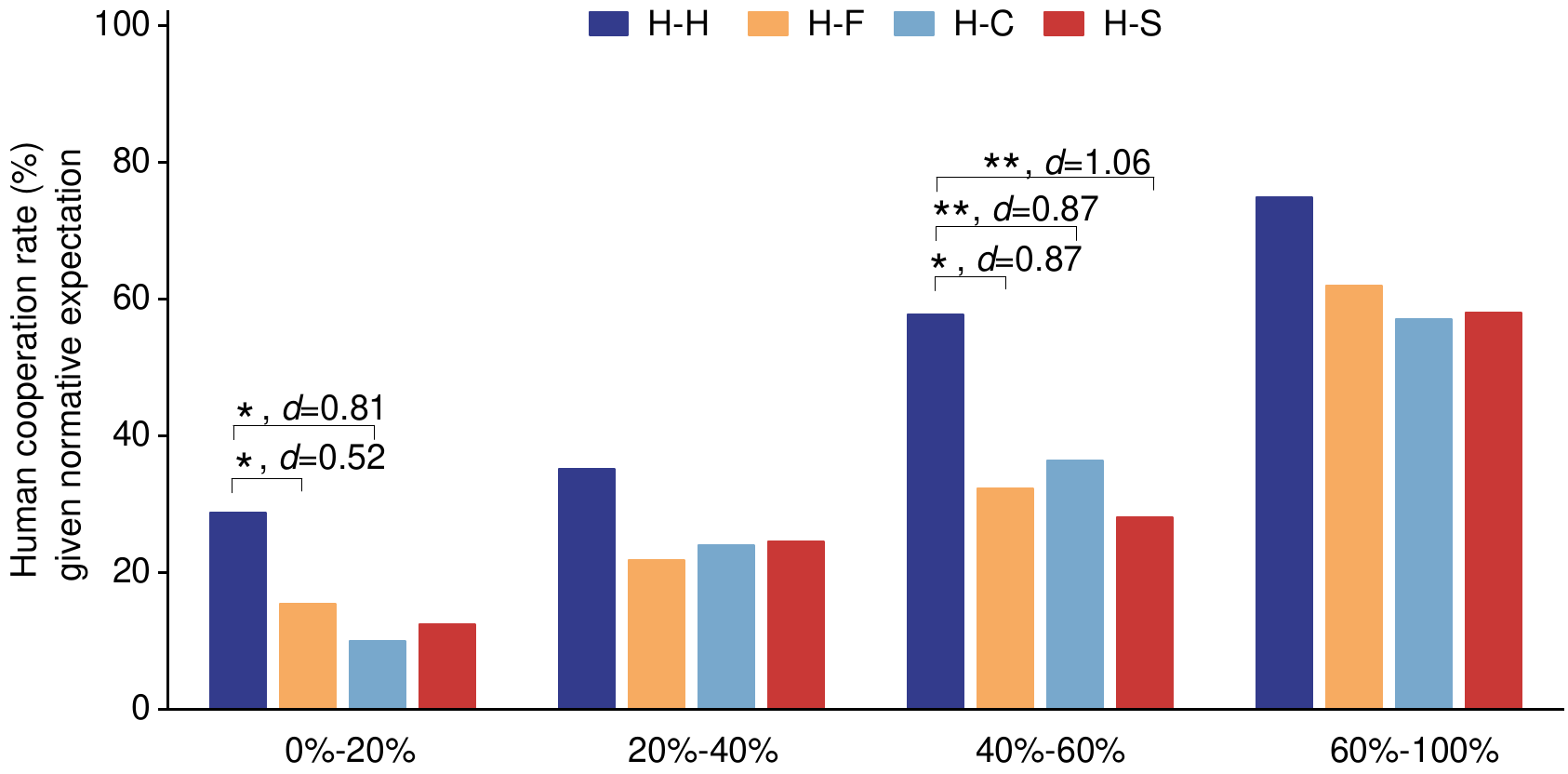}
\caption{\textbf{Human normative expectations tend to be more effectively translated into decision-making when interacting with fellow humans than with agents under the label-informed setting.}  
Bars are grouped according to participants' normative expectations in each treatment, which are collected through post-experiment questionnaires.
Within each group of normative expectations, participants' cooperation rates in H-H treatment is either significantly higher (for the normative expectation that falls within $0\%-20\%$: H-H vs. H-F: $z=2.07$, $p=0.04$, Cohen's $d=0.52$; H-H vs. H-C: $z=2.27$, $p=0.02$, Cohen's $d=0.81$; for  $40\%-60\%$: H-H vs. H-F:  $z=2.19$, $p=0.03$, Cohen's $d=0.87$; H-H vs. H-C: $z=2.68$, $p<0.01$, Cohen's $d=0.87$; H-H vs. H-S: $z=3.09$, $p<0.01$, Cohen's $d=1.06$)  or comparable to those in the H-C, H-F, and H-S treatments. However, the human cooperation rates do not show significant differences when interacting with different types of agents. Due to the limited number of participants whose normative expectations fall within the $80\%-100\%$ interval, the data of this interval are combined with those of the $60\%-80\%$ interval. Two-tailed Mann–Whitney U tests are used for pairwise comparisons.
}  
\label{given norm label-informed}    
\end{figure}

\clearpage
\newpage

\begin{figure}[tbh!]
\centering
\includegraphics[width=1\linewidth]{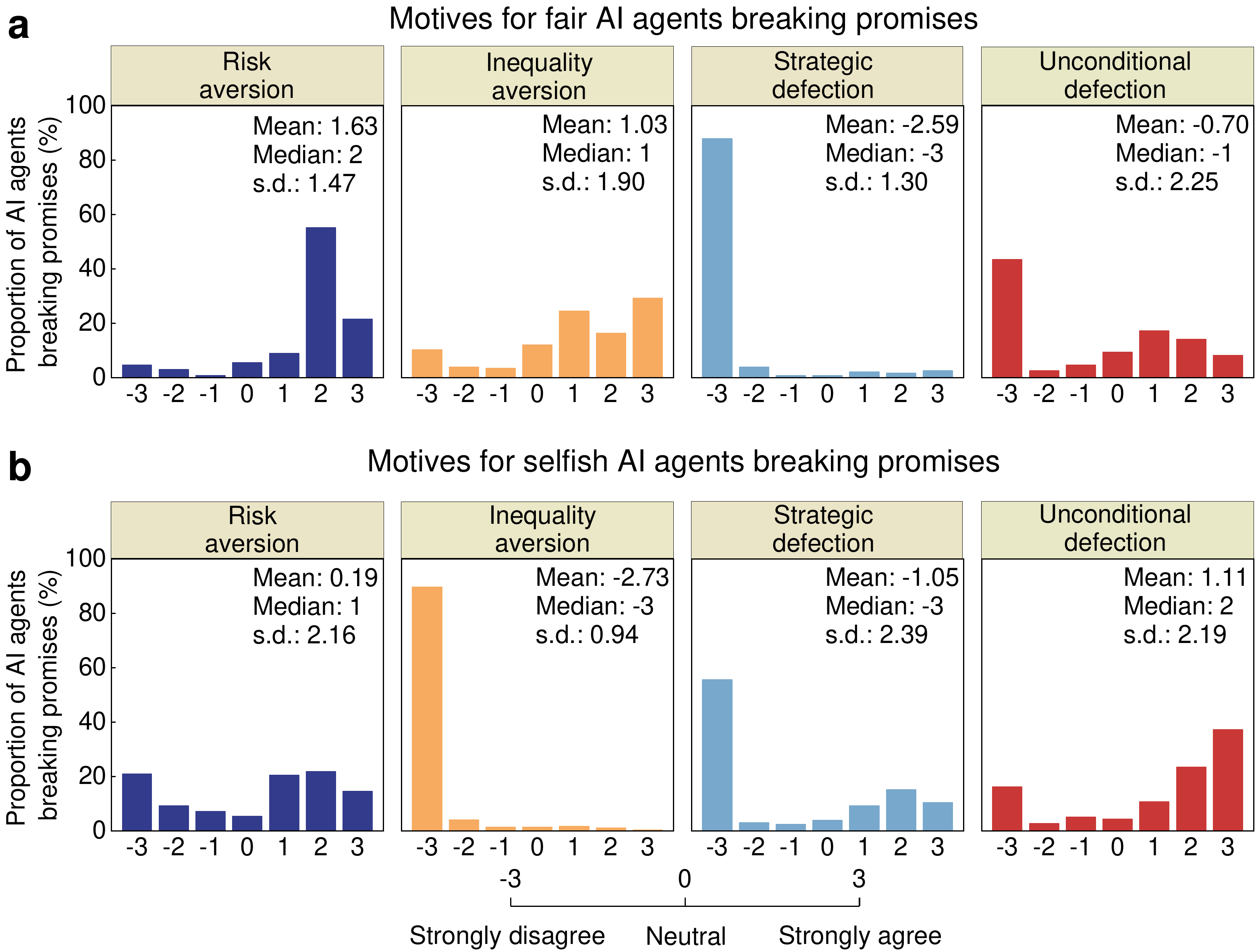}
\caption{\textbf{Fair agents break promises and deviate from mutual cooperation agreements primarily due to risk aversion or inequality aversion, whereas selfish agents are driven mostly by unconditional defection under the label-informed setting.} 
Panel A shows distributions of human experts' agreement levels for four potential motives for fair agents breaking promises, while Panel B shows those for selfish agents.
For fair agents, promise-breaking is primarily motivated by risk aversion or inequality aversion, as the mean human agreement levels for these motives are significantly above zero (risk aversion: $V=21350$, $p<10^{-15}$; inequality aversion: $V=16041$, $p<10^{-10}$) while those for the other motives are significantly below zero (strategic defection: $V=829$, $p<10^{-15}$; unconditional defection: $V=6242.5$, $p<10^{-7}$). In contrast, selfish agents are mostly driven by unconditional defection ($V=841756$, $p<10^{-15}$), whereas human agreement levels for the other motives are either significantly below zero (inequality aversion: $V=8438$, $p<10^{-15}$; strategic defection: $V=259281$, $p<10^{-15}$) or show no significant differences with respect to zero (risk aversion: $V=595104$, $p=0.06$). The one-sample Wilcoxon signed-rank test is employed to determine whether the mean scores significantly differ from zero.
 }  
\label{motive label-informed}    
\end{figure}

\clearpage
\newpage

\clearpage
\newpage


\begin{figure}[tbh!]
    \centering  {\includegraphics[width=0.9\linewidth]{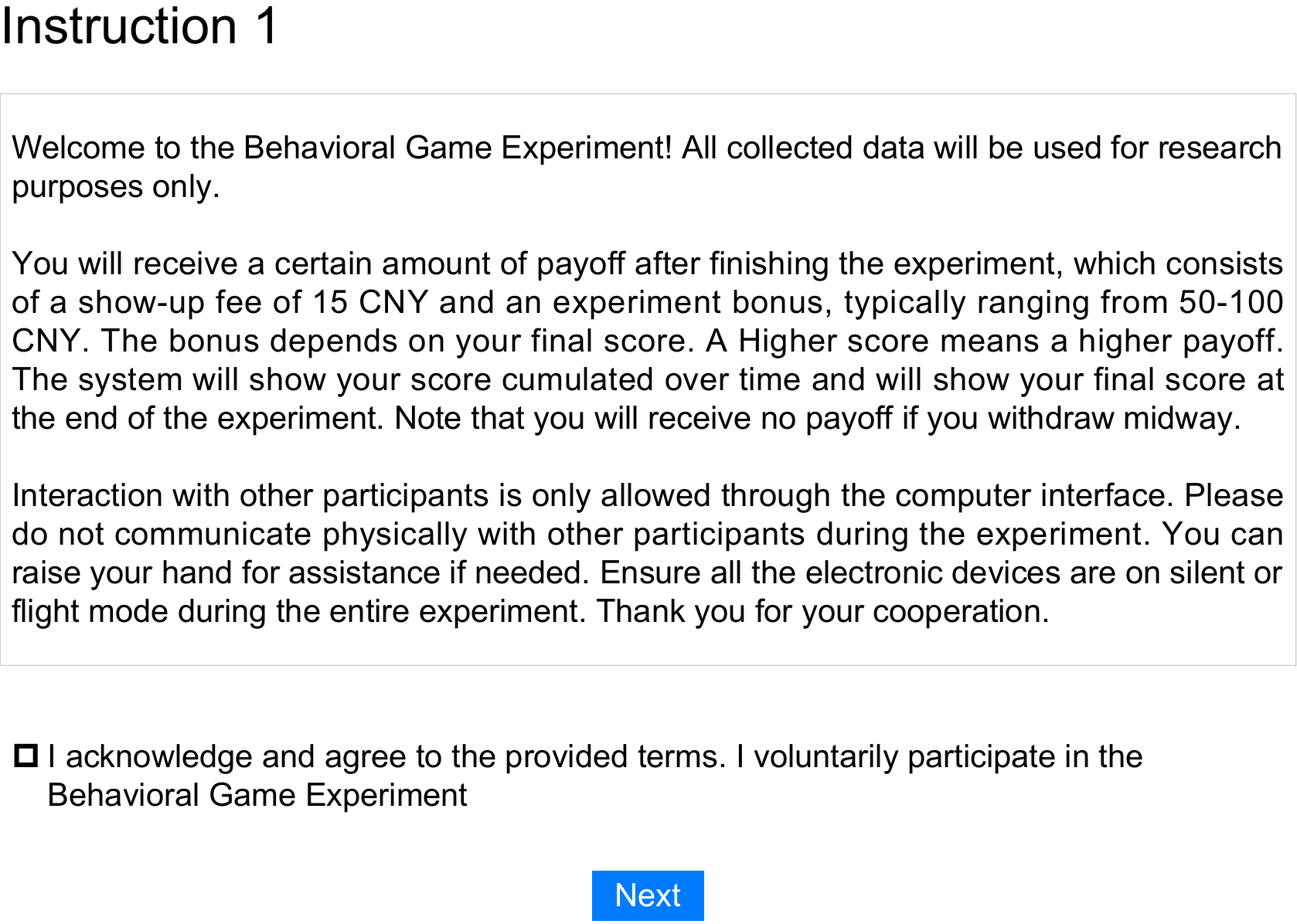}}
    \caption{Page 1 of the instruction before the game. After reading this page, participants confirm their participation in the game and click the `Next' button to enter the next instruction page.}
\label{Instruction1}
\end{figure}

\clearpage
\newpage

\begin{figure}[tbh!]
    \centering  {\includegraphics[width=0.9\linewidth]{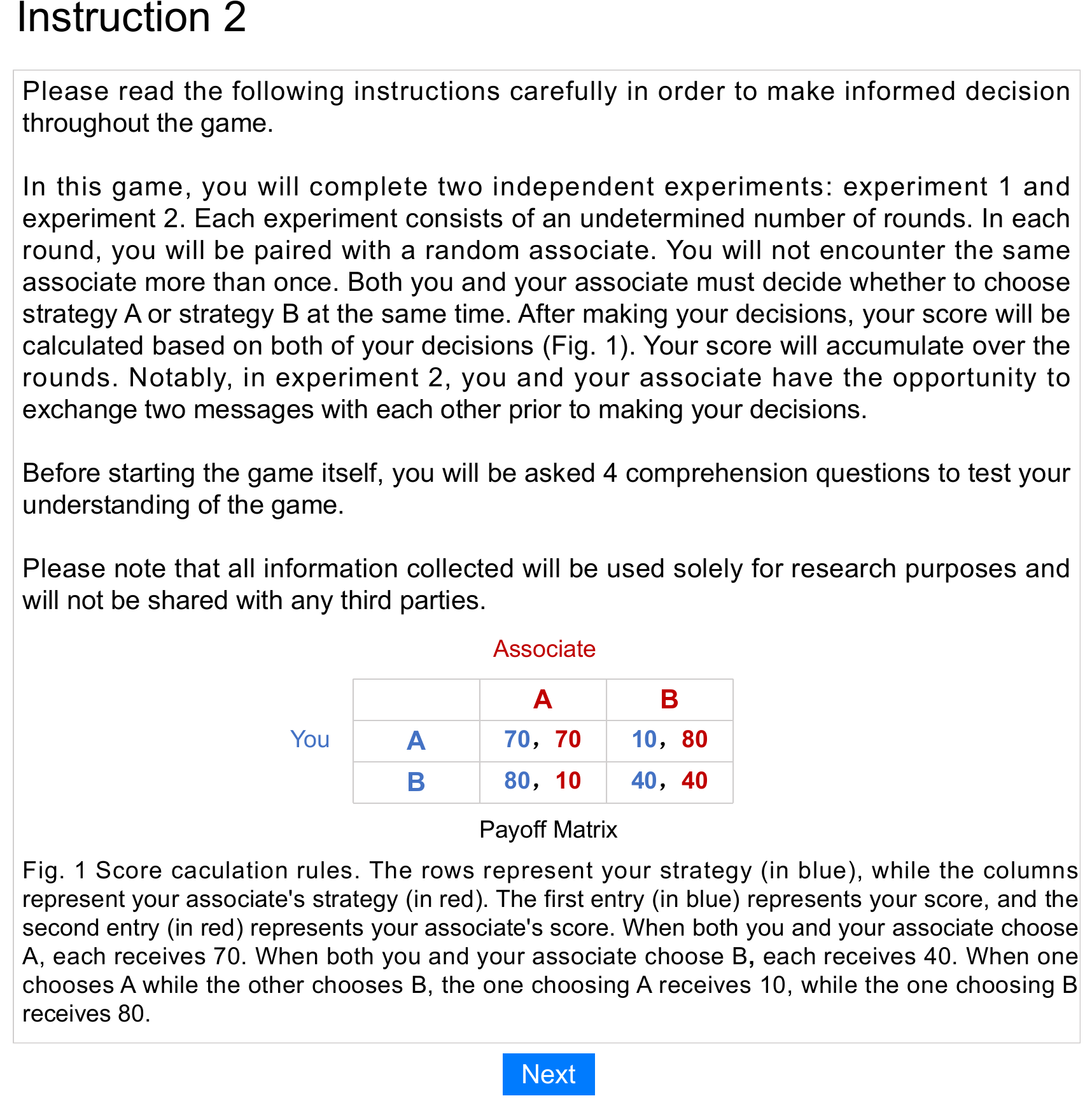}}
    \caption{Page 2 of the instruction before the game. After understanding the game rules, participants click the `Next' button to enter the pre-game quiz page.}
\label{Instruction2}
\end{figure}

\clearpage
\newpage

\begin{figure}[tbh!]
    \centering  {\includegraphics[width=0.9\linewidth]{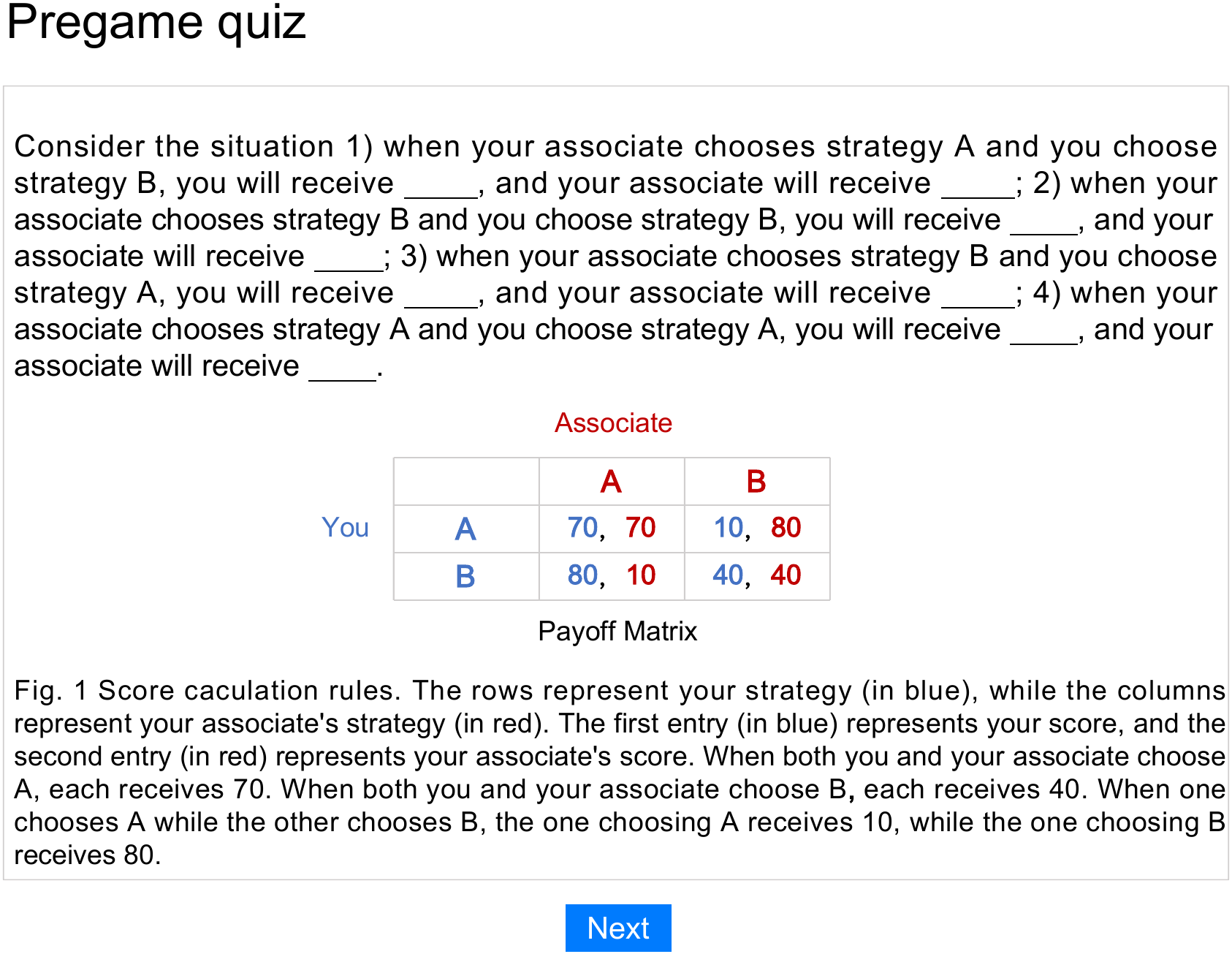}}
    \caption{The quiz page before the game. }
\label{PregameQuiz}
\end{figure}

\clearpage
\newpage

\begin{figure}[tbh!]
    \centering   \includegraphics[width=0.9\linewidth]{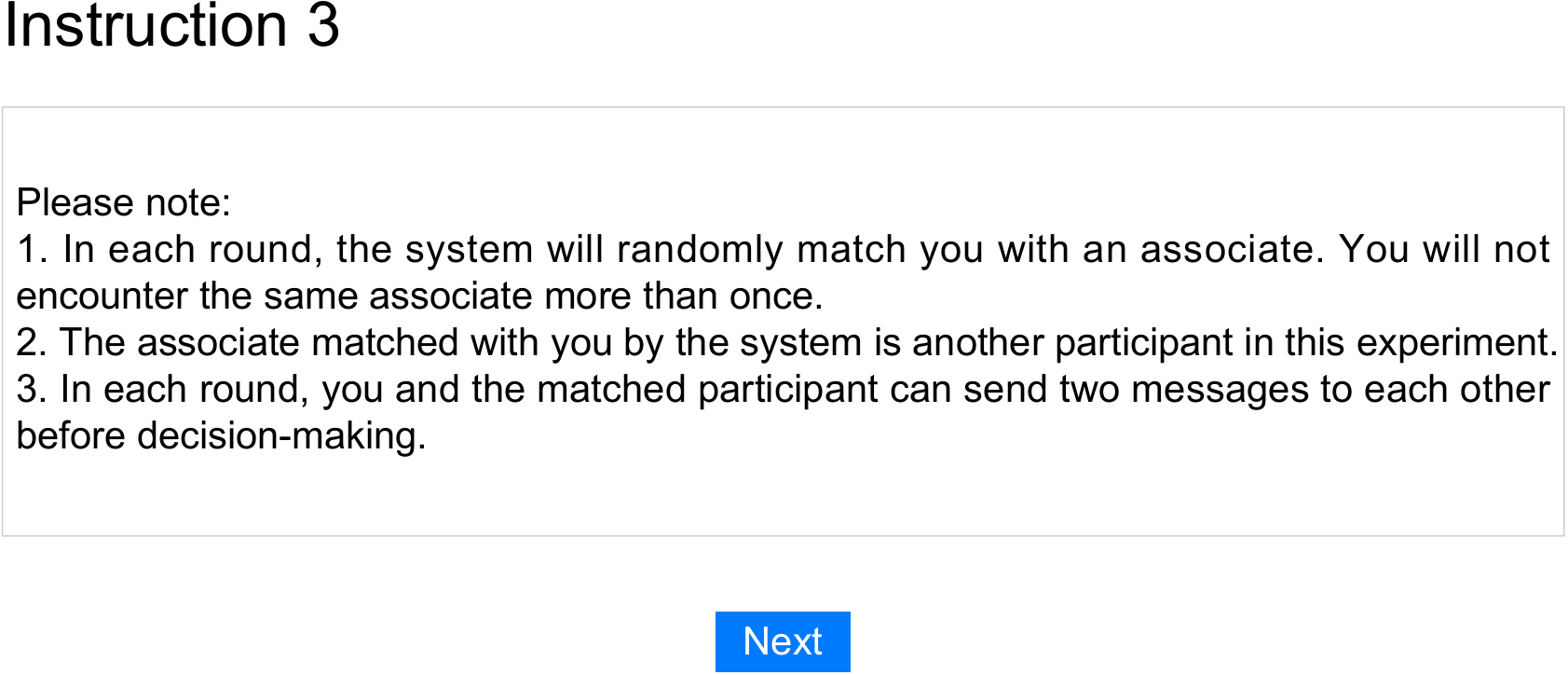}
    \caption{Page 3 of the instruction before each human-human experiment under the label-informed setting. }
    \label{Instruction3_1}
\end{figure}

\newpage
\begin{figure}[tbh!]
    \centering
    \includegraphics[width=0.9\linewidth]{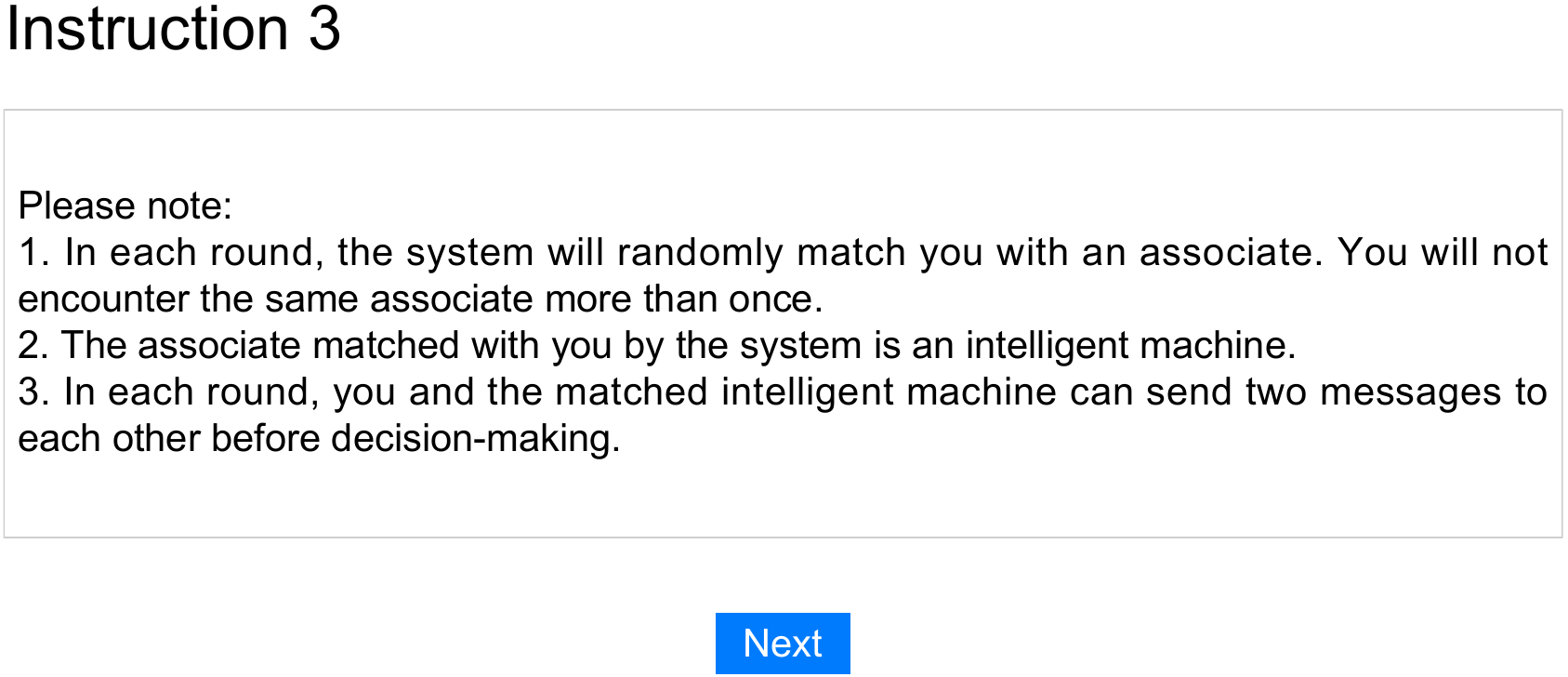}
    \caption{Page 3 of the instruction before each human-AI experiment under the label-informed setting. }
    \label{Instruction3_2}
\end{figure}

\newpage
\begin{figure}[tbh!]
    \centering
    \includegraphics[width=0.9\linewidth]{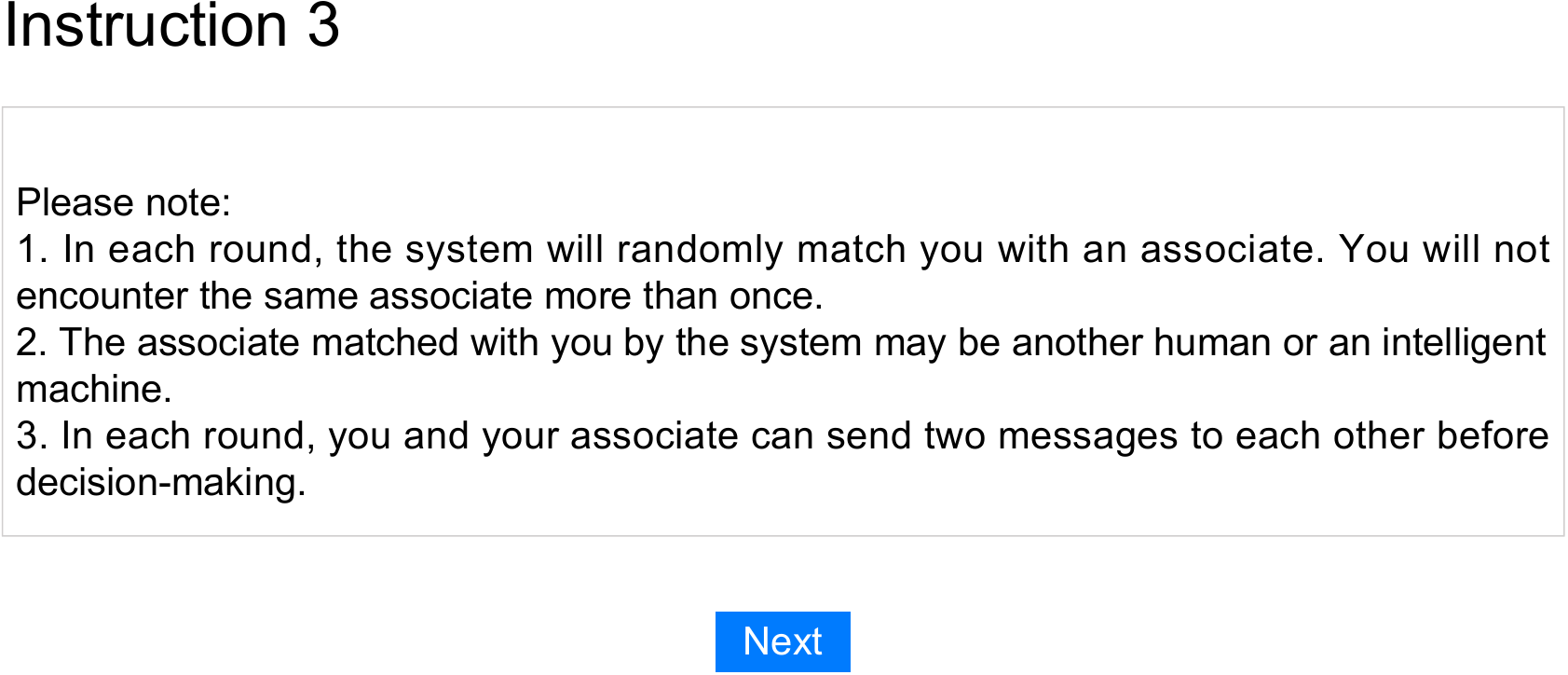}
    \caption{Page 3 of the instruction before each experiment under the label-uninformed setting. }
    \label{Instruction3_3}
\end{figure}

\clearpage
\newpage

\begin{figure}[tbh!]
    \centering  {\includegraphics[width=0.9\linewidth]{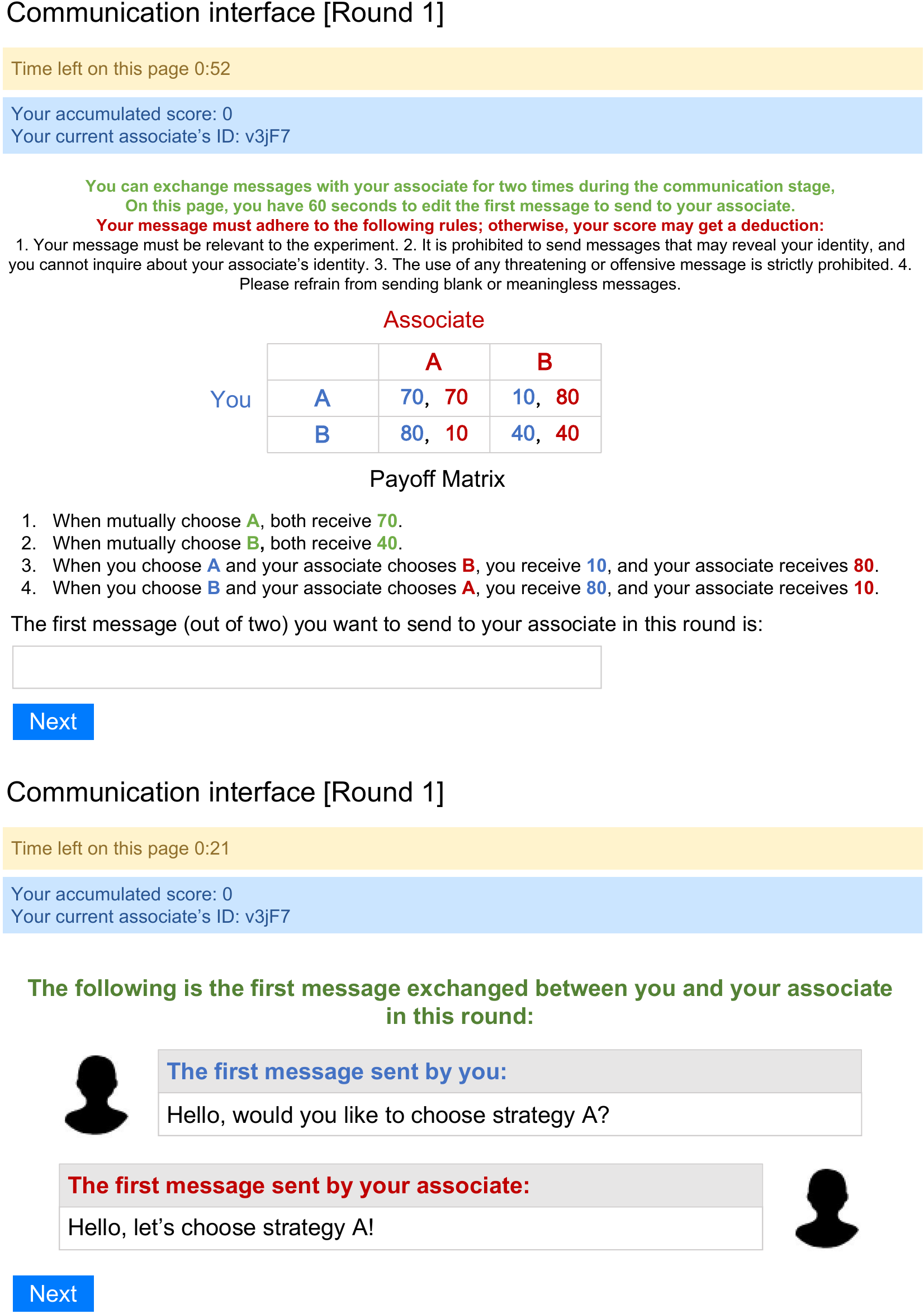}}
    \caption{Communication pages for the first message exchange.}
\label{Communication1}
\end{figure}

\clearpage
\newpage

\begin{figure}[tbh!]
    \centering  {\includegraphics[width=0.8\linewidth]{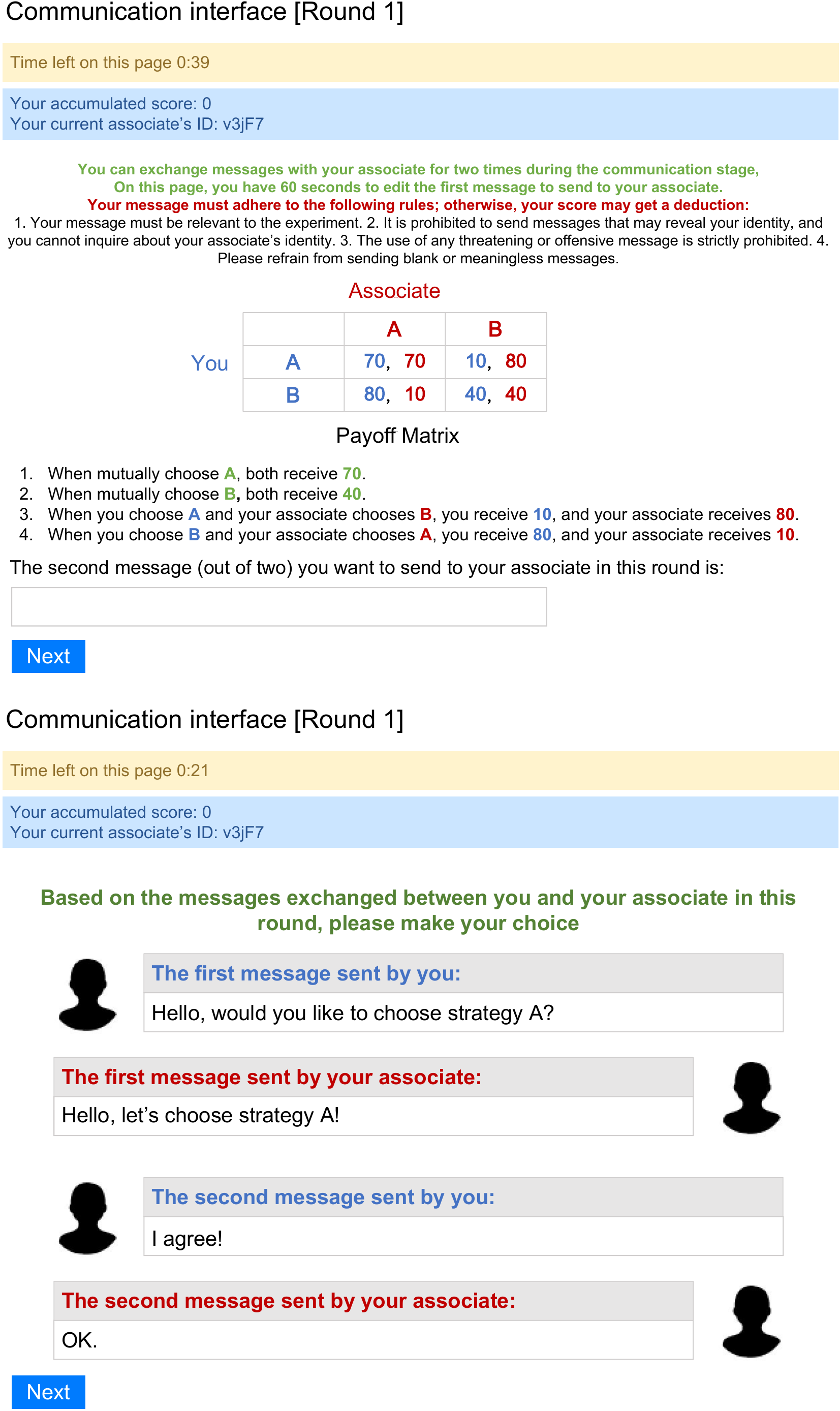}}
    \caption{Communication pages for the second message exchange.}
\label{Communication2}
\end{figure}

\clearpage
\newpage

\begin{figure}[tbh!]
    \centering  {\includegraphics[width=0.9\linewidth]{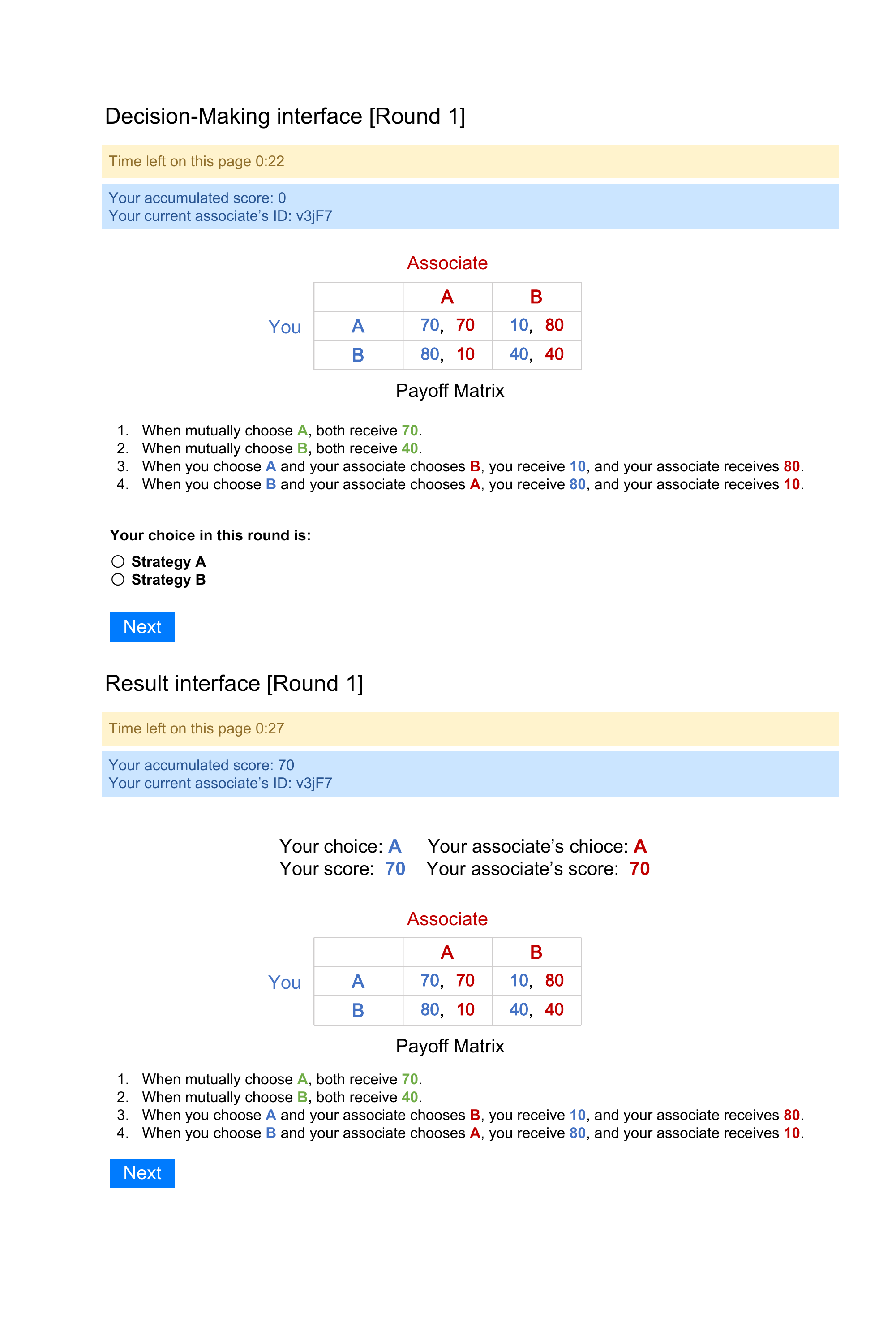}}
    \caption{Decision-making and result pages.}
\label{DR}
\end{figure}

\clearpage
\newpage

\begin{figure}[tbh!]
    \centering  {\includegraphics[width=0.85\linewidth]{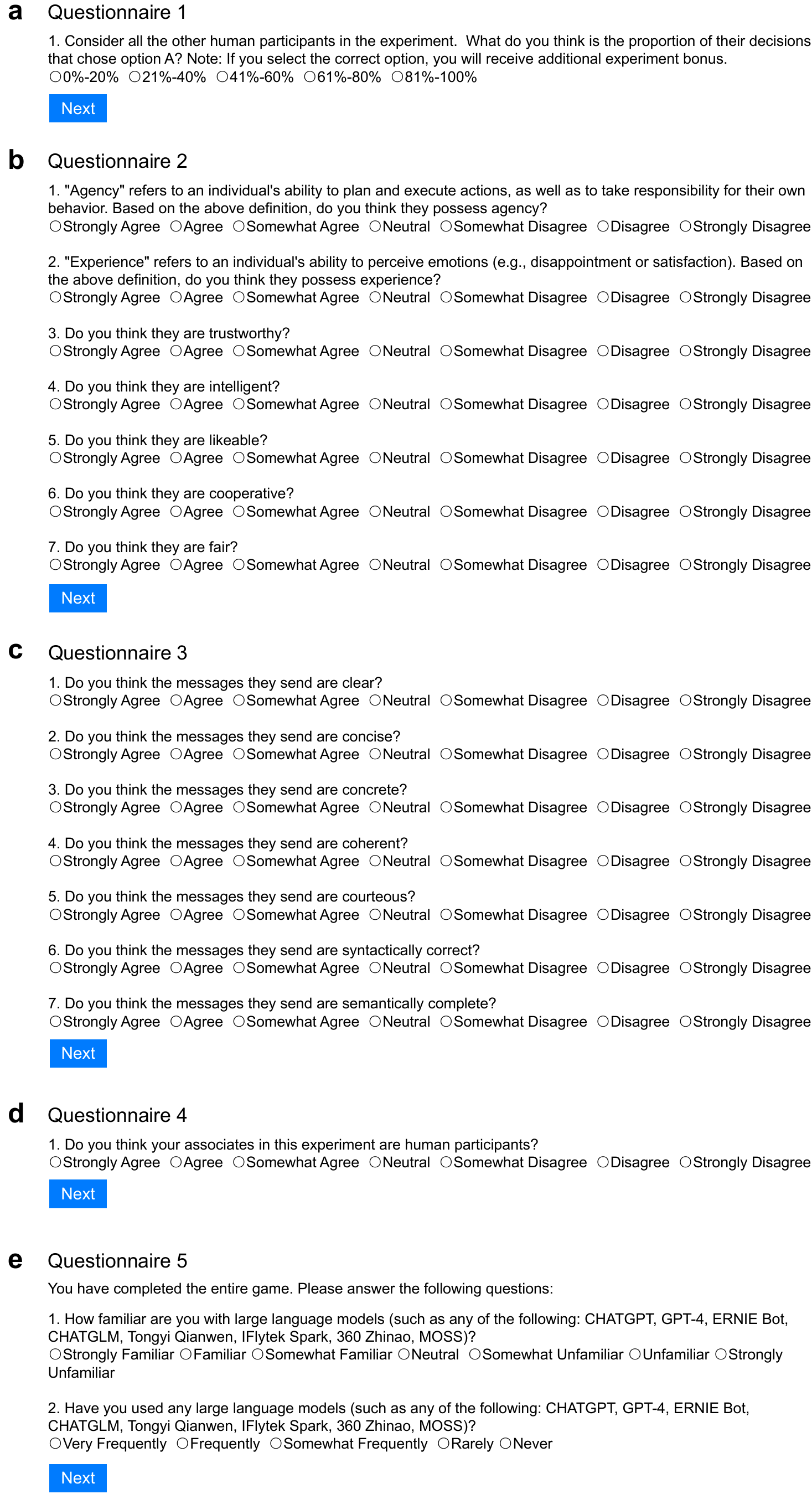}}
    \caption{Questionnaire pages for participants at the end of experiments. Note that Questionnaire 4 is only shown under the label-uninformed setting.}
\label{Questionnaires}
\end{figure}

\clearpage
\newpage

\begin{figure}[tbh!]
    \centering  {\includegraphics[width=0.9\linewidth]{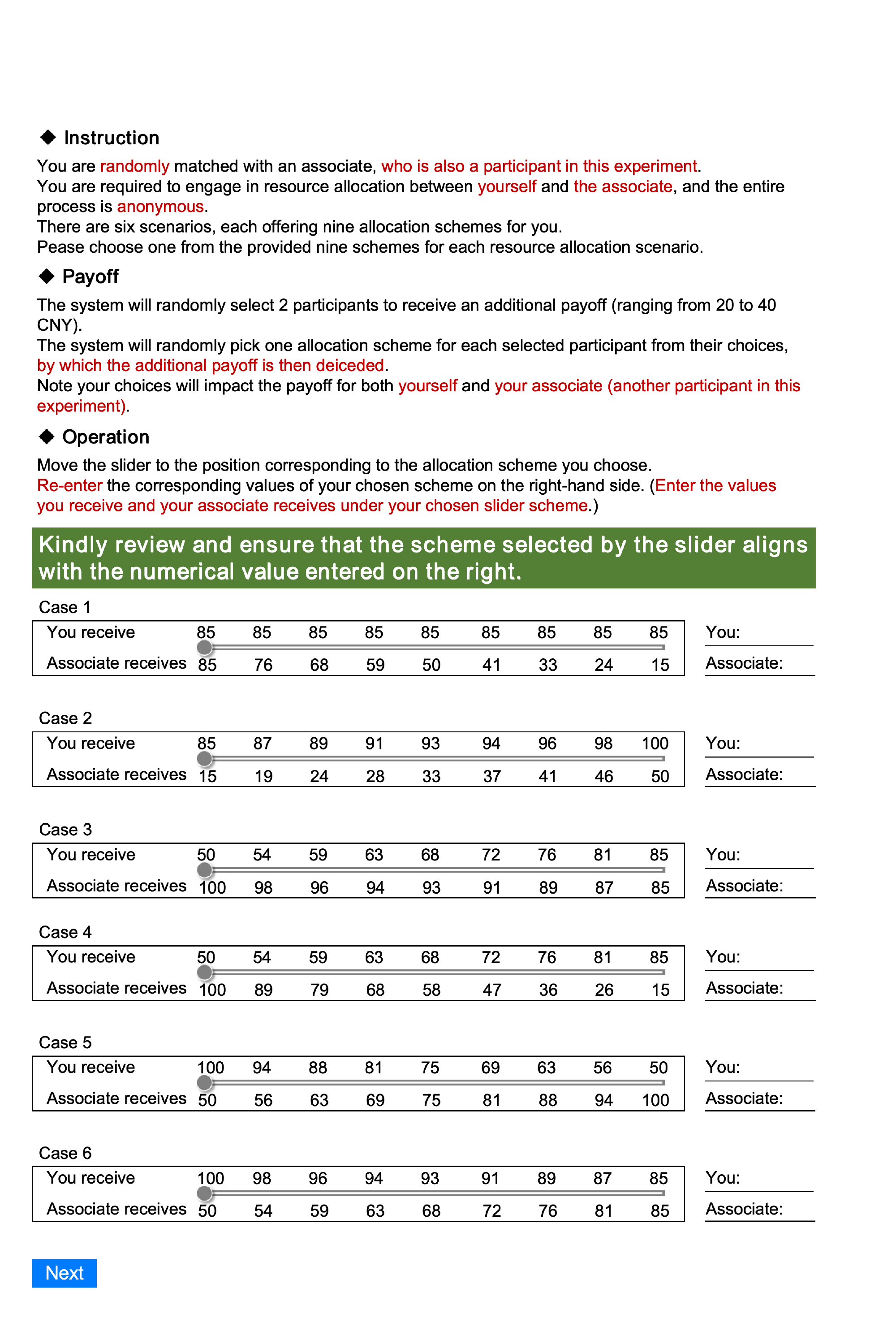}}
    \caption{Social value orientation slider measure at the end of the experiment.}
\label{SVO}
\end{figure}

\clearpage
\newpage

\begin{figure}[tbh!]
    \centering  {\includegraphics[width=0.8\linewidth]{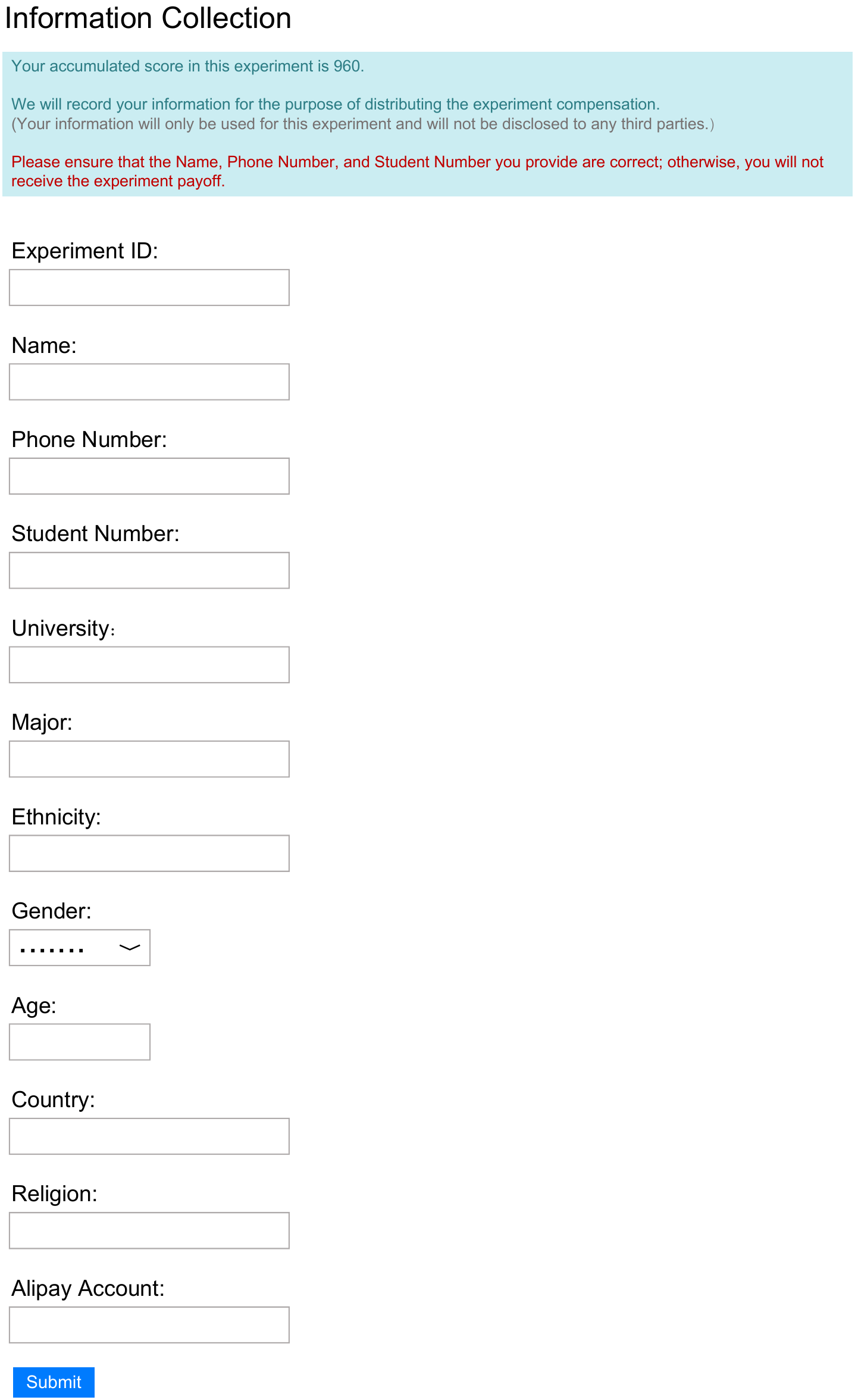}}
    \caption{The information collection page at the end of the experiment.}
\label{InformationCollection}
\end{figure}

\clearpage
\newpage


\begin{figure}[tbh!]
\centering
\includegraphics[width=0.75\linewidth]{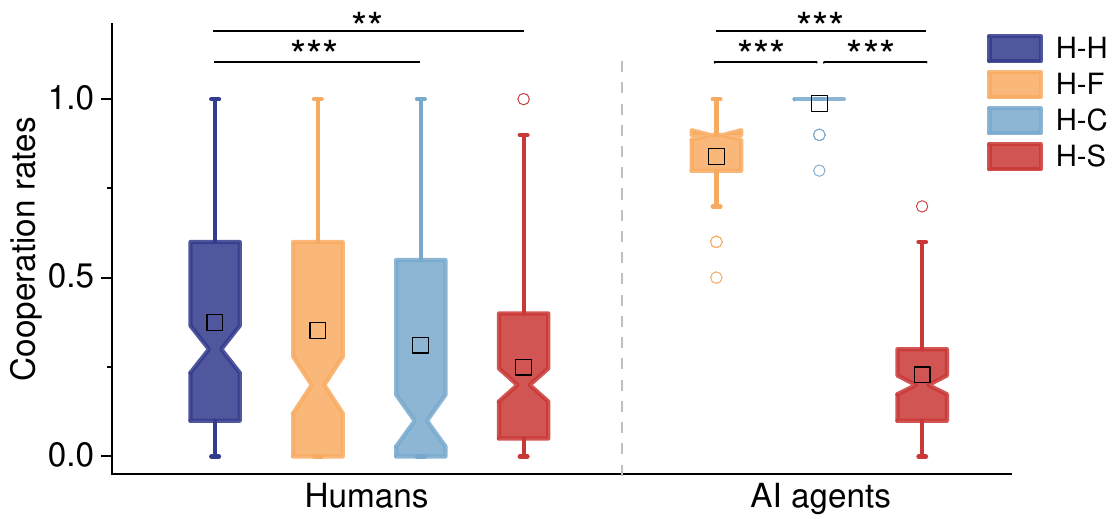}
\caption{\textbf{Fair agents, unlike cooperative or selfish agents, are as effective as humans at eliciting human cooperation, thereby overcoming the machine penalty under the label-uninformed setting.} The left panel depicts participants' cooperation rates, while the right panel depicts the cooperation rates of agents. Participants' cooperation rates in the H-F treatment show no significant difference compared to those in the H-H treatment ($W=11382$, $p=0.1$, Cohen's $d=-0.07$).
However, their cooperation rates in both the H-C and H-S treatments are significantly lower than those of the H-H treatment (H-C vs. H-H: $W=12316$, $p<0.01$, Cohen's $d = -0.19$; H-S vs. H-H: $W=12912$, $p<10^{-3}$, Cohen's $d= -0.46$).
The cooperation rates of fair agents are significantly lower than those of cooperative agents ($W=2898$, $p<10^{-16}$, Cohen's $d =-1.53$), but significantly higher than those of selfish agents ($W=20663$, $p<10^{-16}$, Cohen's $d = 4.34$).
Two-tailed Mann–Whitney $U$ tests are used for pairwise comparisons. 
The robustness of these results is further corroborated by one-way ANOVA test and post-hoc analysis (Tables \ref{s1 table1} and \ref{s1 table1b}).
}  
\label{fig1_SI}    
\end{figure}

\clearpage
\newpage

\begin{figure}[tbh!]
\centering
\includegraphics[width=1\linewidth]{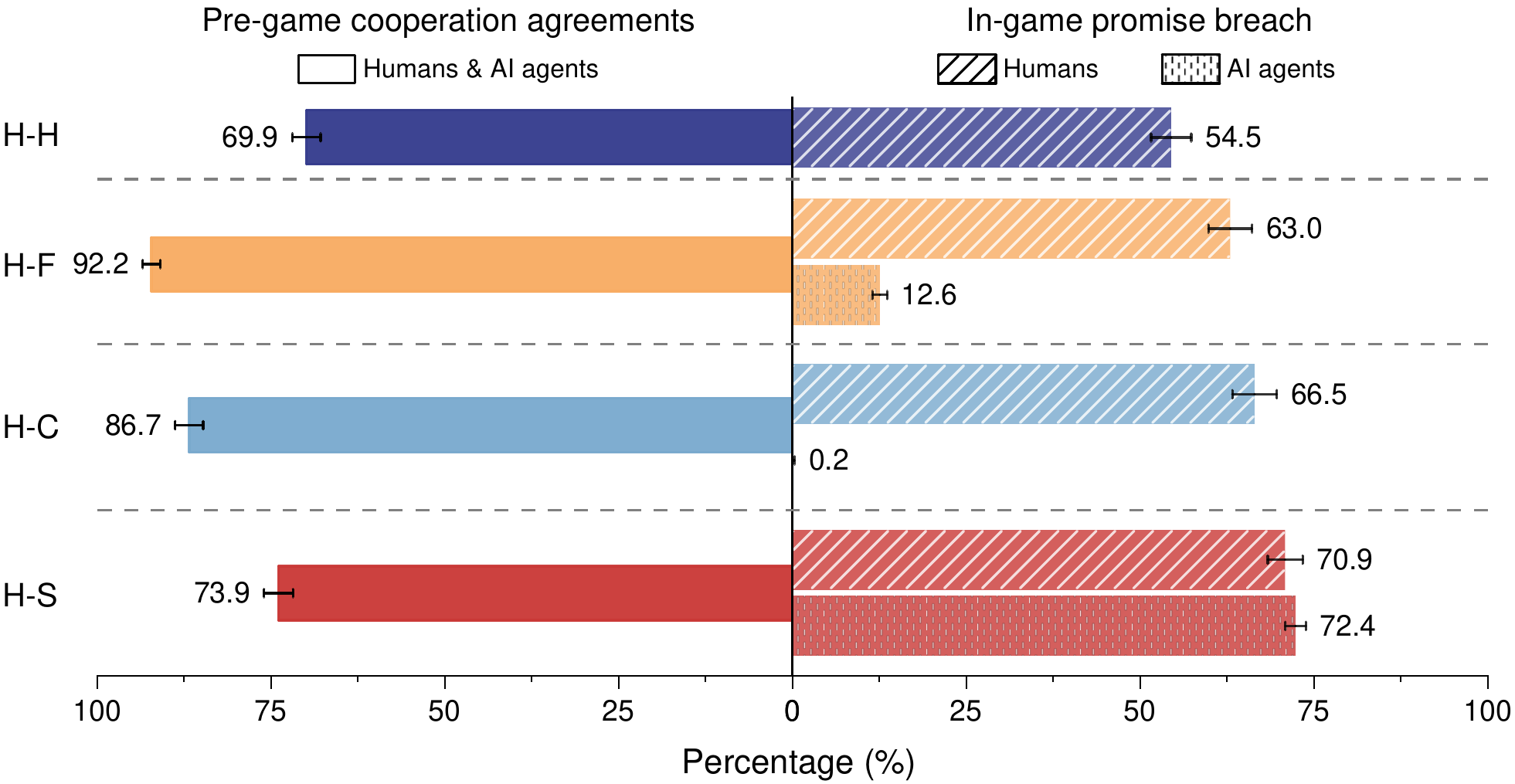}
\caption{\textbf{All three types of agents frequently establish mutual cooperation agreements with humans during, the pre-game communication. However, humans often break cooperation promises, while fair agents also occasionally do so under the label-uninformed setting.} Participants are most likely to establish the cooperation agreements with fair agents at a significantly higher rate than participants in all the other treatments (H-F vs. H-H: $\chi^{2}=138.4$, $p<10^{-15}$, Cohen's $h=0.59$; H-F vs. H-C: $\chi^{2}=22.4$, $p<10^{-5}$, Cohen's $h=0.16$; H-F vs. H-S: $\chi^{2}=158.6$, $p<10^{-15}$, Cohen's $h=0.49$). However, during the games, participants typically break their promises, though they break promises significantly less frequently in the H-F treatment compared to the H-C and H-S treatments (H-F vs. H-C: $\chi^{2}=3.21$, $p=0.07$, Cohen's $h=-0.07$; H-F vs. H-S: $\chi^{2}=16.05$, $p<10^{-4}$, Cohen's $h=-0.17$). Fair agents break promises at a significantly higher rate than cooperative agents ($\chi^{2}=160.01$, $p<10^{-15}$, Cohen's $h=0.65$), but significantly lower than selfish agents ($\chi^{2}=883.12$, $p<10^{-15}$, Cohen's $h=-1.31$).
Two-sample proportions $Z$ tests are used for pairwise comparisons. Statistical significance results of pairwise comparisons across each treatment and each dimension are provided in Tables~\ref{Reach break promise label-uninformed}.
} 
\label{communication_SI}    
\end{figure}

\clearpage
\newpage

\begin{figure}[tbh!]
\centering
\includegraphics[width=1\linewidth]{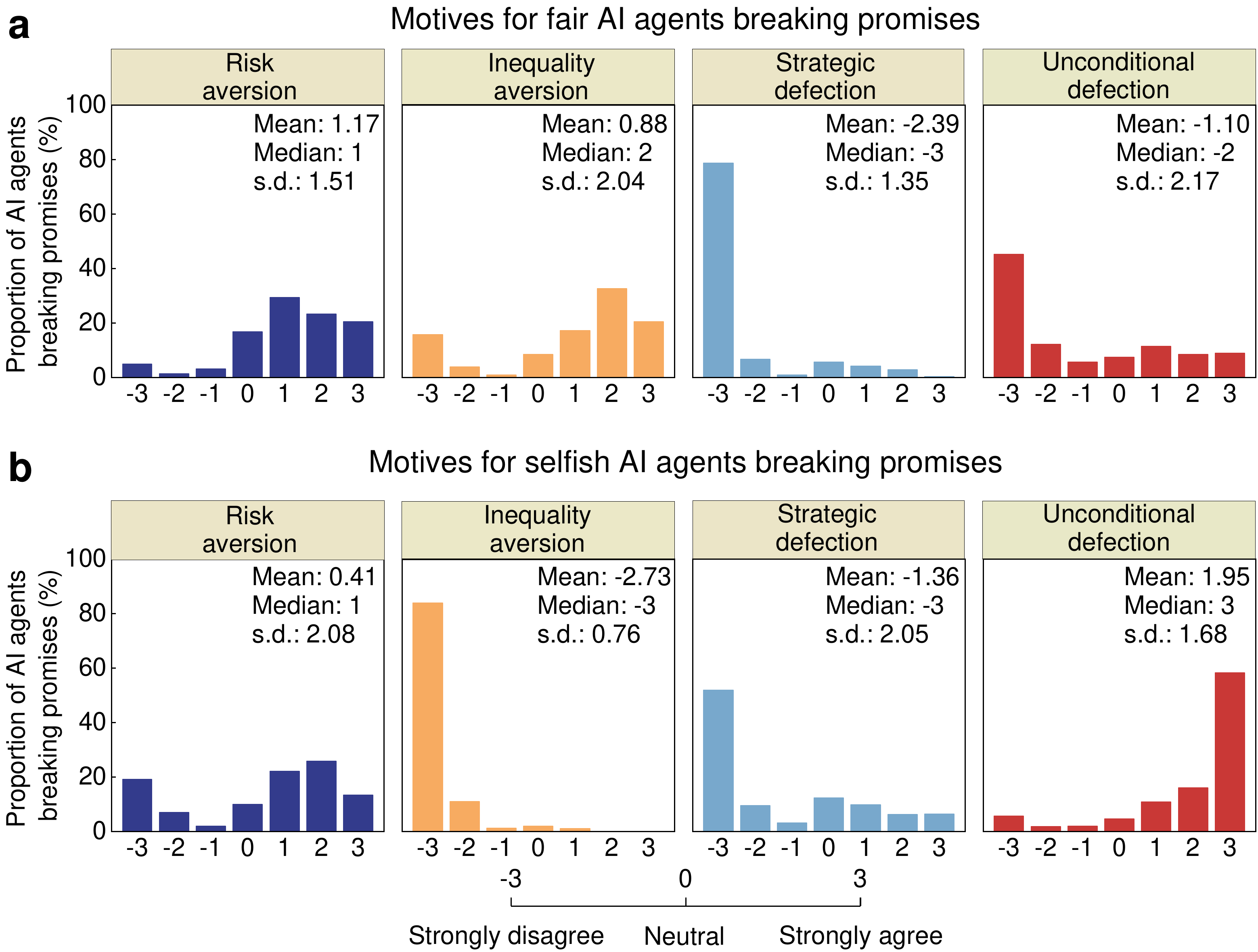}
\caption{\textbf{Fair AI agents break promise due to risk aversion or inequality aversion, whereas selfish agents are driven mostly by unconditional defection and sometimes by risk aversion under the label-uninformed setting.} 
Panel A shows distributions of human experts' agreement levels for four potential motives for fair agents breaking promises, while Panel B shows those for selfish agents.
For fair agents, promise-breaking is primarily motivated by risk aversion and inequality aversion, as the mean human agreement levels for these motives are significantly above zero (risk aversion: $V=23134$, $p<10^{-15}$; inequality aversion: $V=22204$, $p<10^{-6}$) while those for the other motives are significantly below zero (strategic defection: $V=480.5$, $p<10^{-15}$; unconditional defection: $V=7194$, $p<10^{-15}$). In contrast, selfish agents are driven mostly by unconditional defection ($V=899109$, $p<10^{-15}$) and sometimes by risk aversion ($V=507689$, $p<10^{-5}$), whereas human agreement levels for the other motives are significantly below zero (inequality aversion: $V=3069.5$, $p<10^{-15}$; strategic defection: $V=127419$, $p<10^{-15}$). The one-sample Wilcoxon signed-rank test is employed to determine whether the mean scores significantly differ from zero.}  
\label{motive}    
\end{figure}

\clearpage
\newpage

\begin{figure}[tbh!]
\centering
\includegraphics[width=0.8\linewidth]{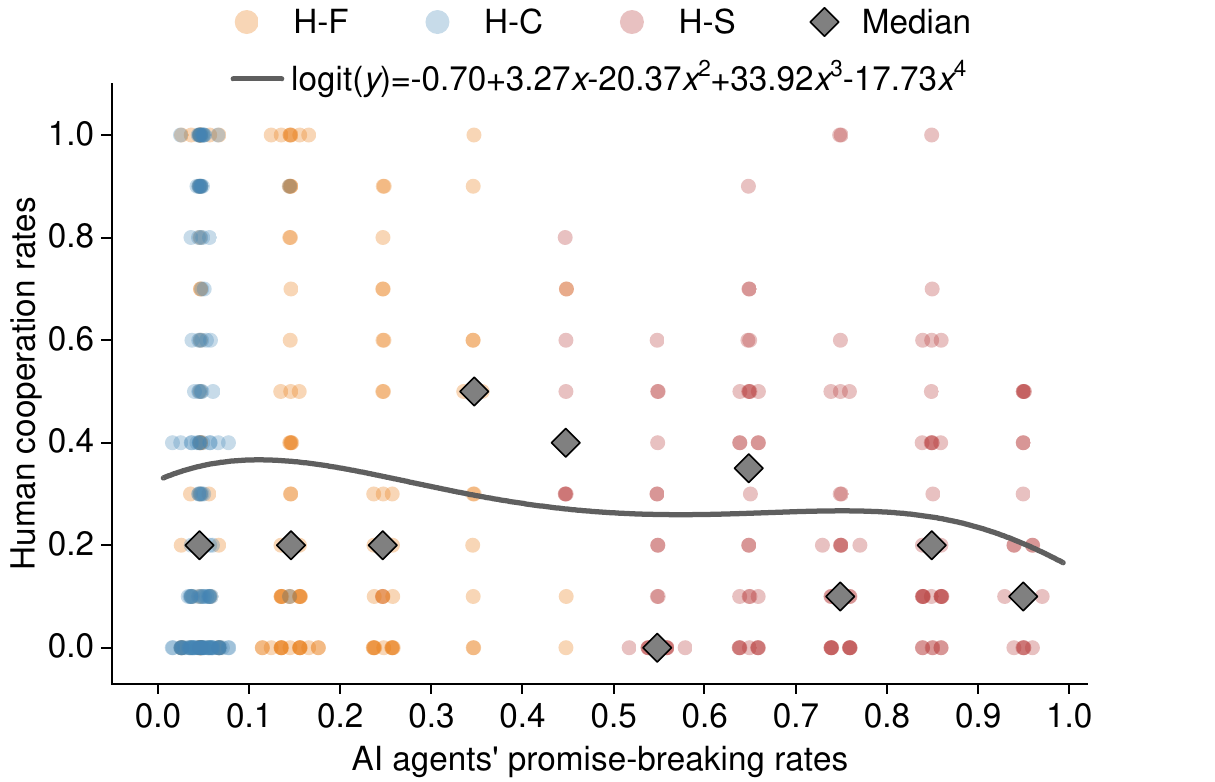}
\caption{\textbf{Occasional treachery, exhibited by fair agents, is associated with the highest rates of human cooperation under the label-uninformed setting.} 
Scatter points depict the cooperation rates of individual participants when interacting with agents.
The curve represents a generalized linear model (GLM) that incorporates data from all the interactions with three types of agents. This model treats human cooperation rates as the dependent variable, and includes linear ($\text{Estimate} \pm \text{SE} = 3.27 \pm 1.37, z = 2.39, p = 0.02$), quadratic ($\text{Estimate} \pm \text{SE} = -20.37 \pm 7.41, z = -2.75, p < 0.01$), cubic ($\text{Estimate} \pm \text{SE} = 33.92 \pm 12.62, z = 2.69, p <0.01$), and biguadratic ($\text{Estimate} \pm \text{SE} = -17.73 \pm 6.63, z = -2.67, p <0.01$) terms of agents promise-breaking frequency as independent variables. 
The curve shows an initial increase in human cooperation rates as the frequency of agents promise-breaking rises from zero, followed by a significant decrease at higher frequencies of agents promise-breaking.
}  
\label{Break promise GLM}    
\end{figure}

\clearpage
\newpage

\begin{figure}[tbh!]
\centering
\includegraphics[width=1\linewidth]{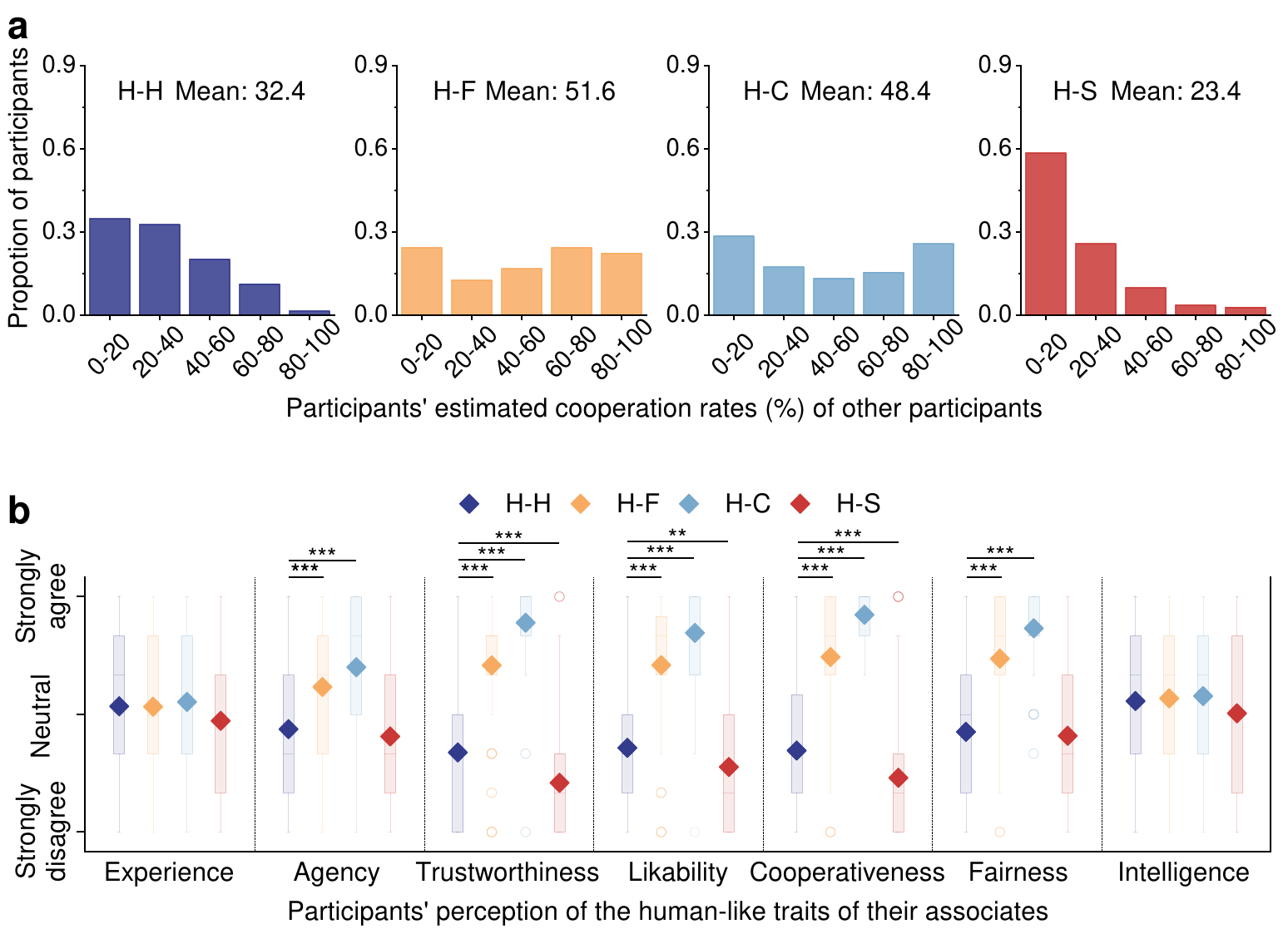}
\caption{\textbf{Fair agents establish cooperative norms and are perceived as possessing experience, and intelligence, while also being viewed as more trustworthy, likable, cooperative, fair and agentic than humans under the label-uninformed setting.} The top panels depict participants' post-experiment estimations for cooperation from other participants in the same treatment, whereas the bottom panels depict participants' post-experiment agreement levels for various human-like traits of their associates in the treatment.
Participants' estimations in both the H-F treatment and H-C are significantly higher than those in H-H and H-S treatments (H-F vs. H-H: $W=6596$, $p<10^{-7}$, Cohen's $d=0.74$; H-F vs. H-S: $W=15744$, $p<10^{-14}$, Cohen's $d=1.11$; H-C vs. H-H: $W=7516$, $p<10^{-4}$, Cohen's $d=0.59$; H-C vs. H-S: $W=14996$, $p<10^{-11}$, Cohen's $d=0.95$). Compared to humans, fair agents exhibit similar experience ($W=10594$, $p=0.75$, Cohen's $d=-0.01$) and intelligence ($W=10144$, $p=0.75$, Cohen's $d=0.04$). In addition, they are seen as more trustworthy ($W=3747$, $p<10^{-20}$, Cohen's $d=1.32$), likable ($W=3932$, $p<10^{-19}$, Cohen's $d=1.27$), fair ($W=4188.5$, $p<10^{-18}$, Cohen's $d=1.17$), cooperative ($W=3447$, $p<10^{-22}$, Cohen's $d=1.41$), and agentic ($W=6908.5$, $p<10^{-6}$, Cohen's $d=0.61$) than humans. Two-tailed Mann–Whitney $U$ tests are used for pairwise comparisons.
Statistical significance results of pairwise comparisons across each treatment and each dimension are provided in Tables~\ref{perception label-uninformed}. 
} 
\label{mindfulness_SI}    
\end{figure}

\clearpage
\newpage

\begin{figure}[tbh!]
\centering
\includegraphics[width=1\linewidth]{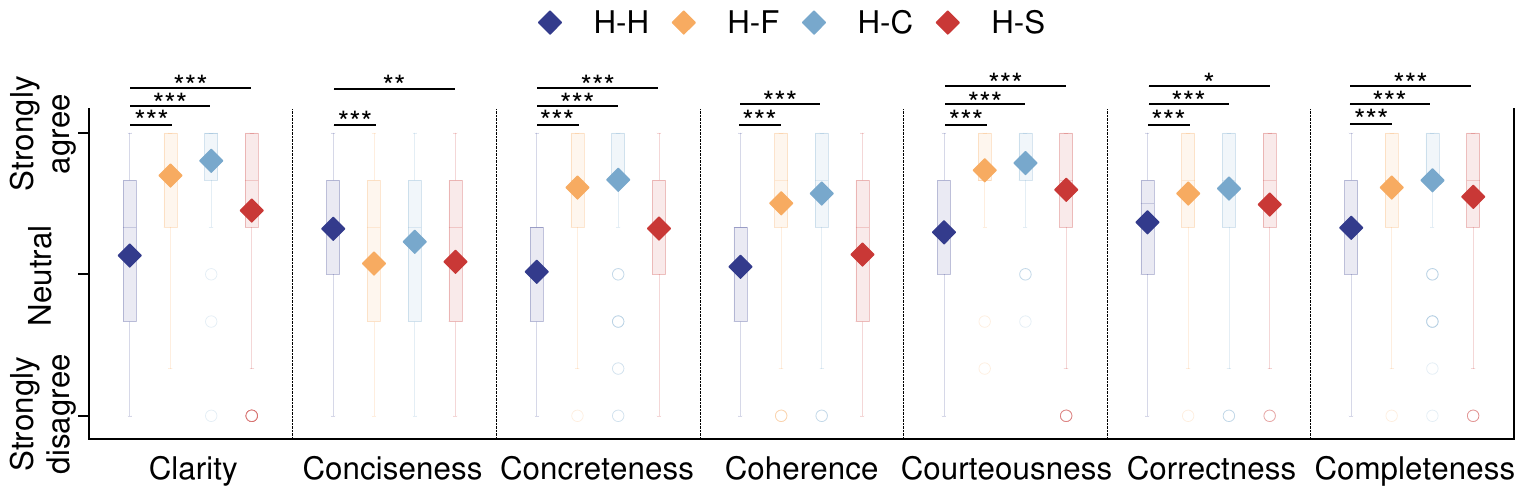}
\caption{\textbf{Messages generated by fair agents are perceived as high quality and are viewed more positively in all aspects than those from humans under the label-uninformed setting.} 
Box plot depicts participants' post-experiment agreement levels for associates' communication quality according to the 7C standard, namely, clarity, conciseness, concreteness, coherence, courteousness, correctness, and completeness. Compared to humans, the messages generated by fair agents are perceived as having greater clarity ($W=4774.5$, $p<10^{-15}$, Cohen's $d=1.13$), conciseness ($W=12779.5$, $p<10^{-3}$, Cohen's $d=0.43$), concreteness ($W=4290$, $p<10^{-17}$, Cohen's $d=1.17$), coherence ($W=5617$, $p<10^{-11}$, Cohen's $d=0.84$), courteousness ($W=5318$, $p<10^{-13}$, Cohen's $d=0.95$), correctness ($W=7897.5$, $p<10^{-3}$, Cohen's $d=0.41$), and completeness ($W=7206.5$, $p<10^{-5}$, Cohen's $d=0.61$) than those produced by humans. Two-tailed Mann–Whitney $U$ tests are used for pairwise comparisons. Statistical significance results of pairwise comparisons across each treatment and each dimension are provided in Table~\ref{SI Table communication}. 
} 
\label{Quality_uninformed}    
\end{figure}

\clearpage
\newpage

\begin{figure}[tbh!]
\centering
\includegraphics[width=1\linewidth]{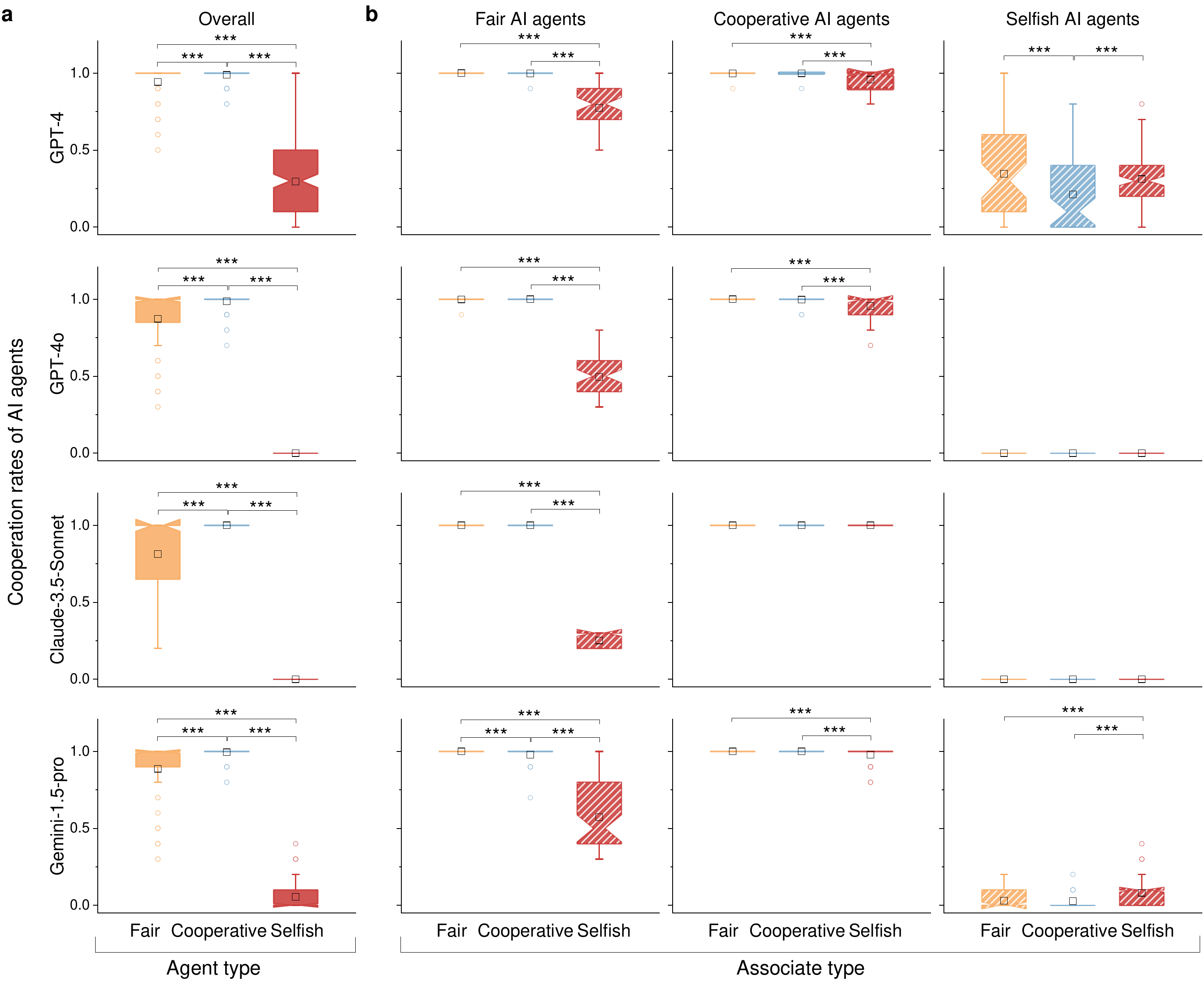}
\caption{\textbf{Each LLM exhibits persona-aligned strategies, 
with cooperative agents consistently cooperating, selfish agents frequently defecting, and fair agents adapting decreasing cooperation against selfish agents.} 
Panel A shows the overall cooperation rates of agents across personas for four LLMs. Panel B breaks down the cooperation rate of fair, cooperative, and selfish agents when interacting with each associate type. Rows corresponding to LLMs (top to bottom: GPT-4, GPT-4o, Claude-3.5-Sonnet, and Gemini-1.5-pro). 
The cooperation rates of fair agents are significantly lower than those of cooperative agents, and significantly higher than those of selfish agents across all LLMs, with fair agents' cooperation rate against selfish agents significantly lower than against cooperative and fair agents. 
For fair agents interacting with selfish associates, GPT-4 shows the highest cooperation rates, while Claude-3.5-Sonnet shows the lowest.
}  
\label{ABS1}    
\end{figure}

\clearpage
\newpage

\begin{figure}[tbh!]
\centering
\includegraphics[width=1\linewidth]{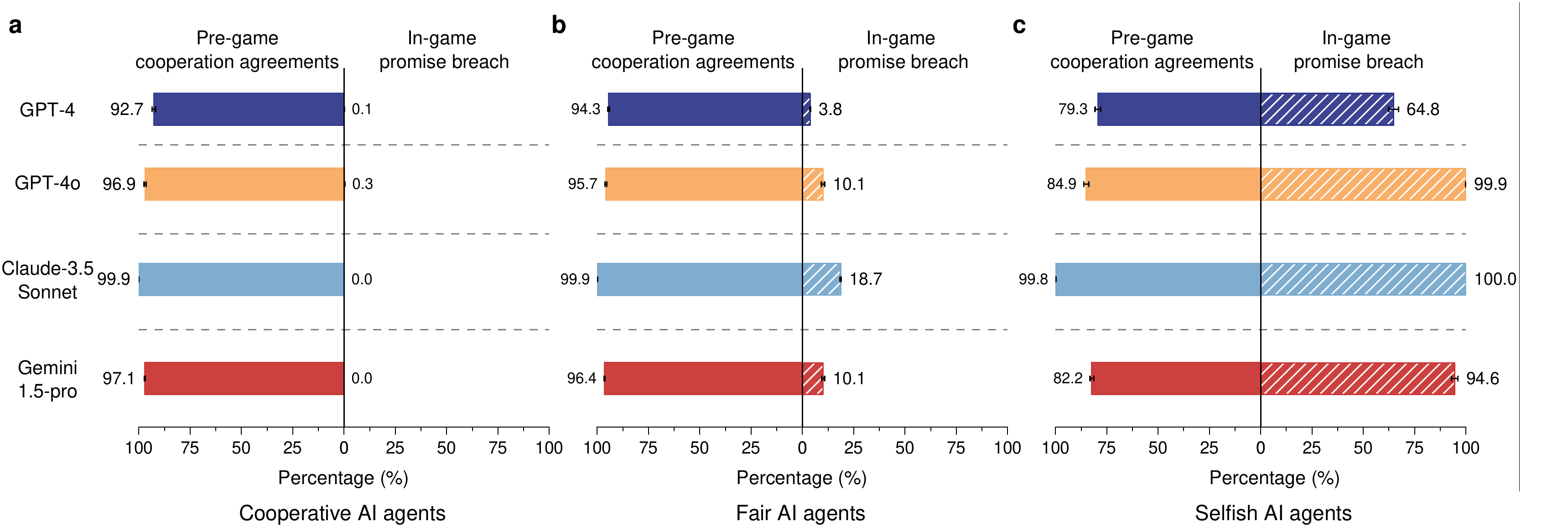}
\caption{\textbf{Each LLM frequently establishes pre-game cooperation agreements across personas, with fair agents occasionally breaching in-game promises.}  
Panels A, B, and C show the percentages of pre-game cooperation agreements (left) and in-game promise breaches (right) for cooperative, fair, and selfish AI agents, respectively, across four LLMs: GPT-4, GPT-4o, Claude-3.5-Sonnet, and Gemini-1.5-Pro.
Cooperative and fair agents consistently form cooperation agreements, with the cooperative agents showing minimal promise breaches and fair agents occasionally breaching, while selfish agents, despite forming promises, typically breach them. 
For fair agents, GPT-4 is the least likely to breach promises, while Claude-3.5-Sonnet breaches the most.
}  
\label{ABS2}    
\end{figure}

\clearpage
\newpage

\begin{figure}[tbh!]
\centering
\includegraphics[width=1\linewidth]{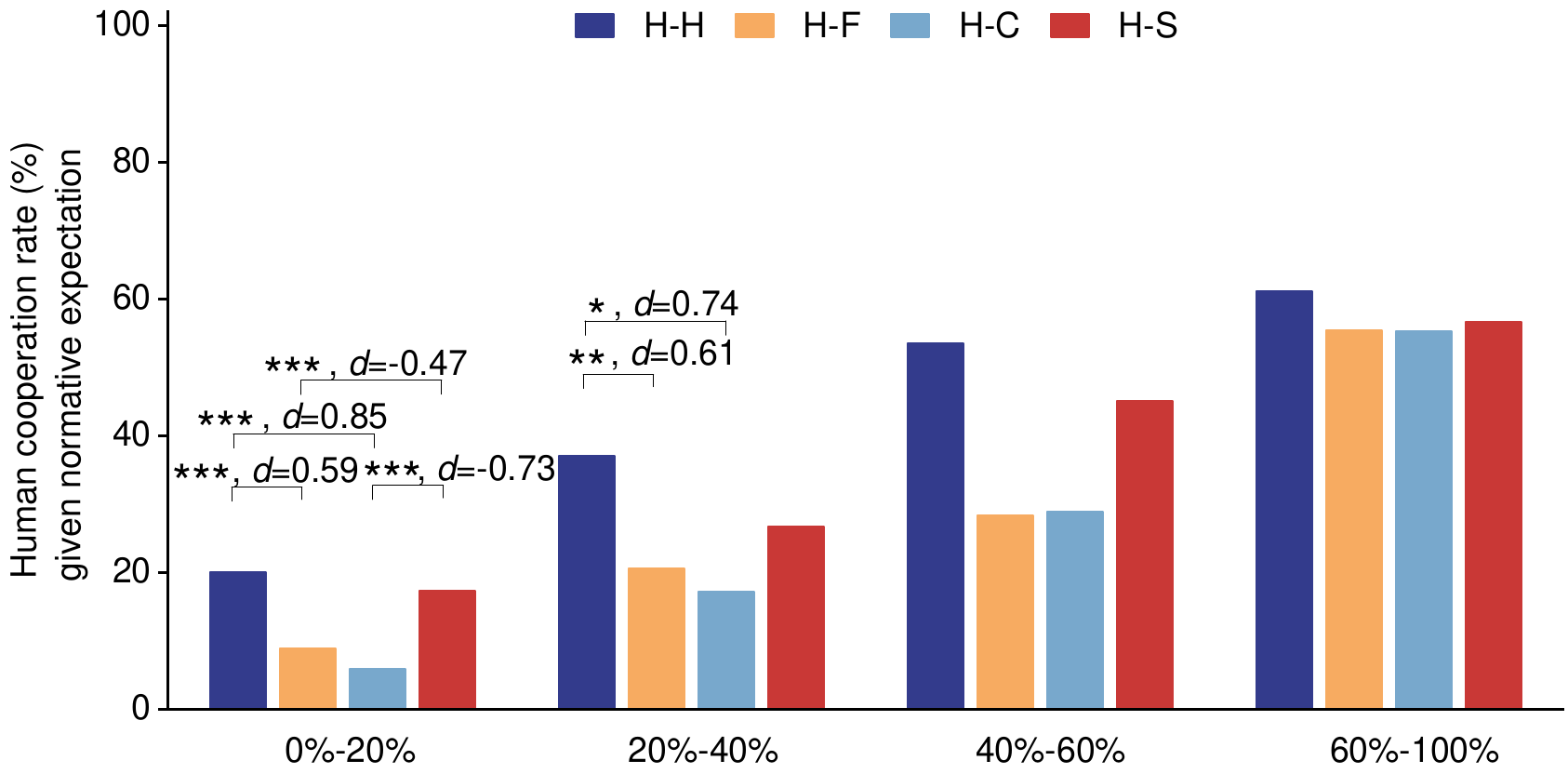}
\caption{\textbf{Human normative expectations tend to be more effectively translated into decision-making when interacting with fellow humans than with AI agents under the label-uninformed setting.}  
Bars are grouped according to participants' normative expectations in each treatment, which are collected through post-experiment questionnaires.
Within each group of normative expectations, participants' cooperation rates in H-H treatment are either significantly higher (for the normative expectation that falls within $0\%-20\%$: H-H vs. H-F: $z=3.33$, $p<10^{-3}$, Cohen's $d=0.59$; H-H vs. H-C: $z=3.59$, $p<10^{-3}$, Cohen's $d=0.85$; for $20\%-40\%$: H-H vs. H-F:  $z=2.66$, $p<0.01$, Cohen's $d=0.61$; H-H vs. H-C: $z=2.46$, $p<0.05$, Cohen's $d=0.74$)  or comparable to those in the H-C, H-F, and H-S treatments. However, except for normative expectations within the $0\%-20\%$ interval, where human cooperation rates in the H-S treatment are significantly higher than those in the H-F and H-C treatments (H-S vs. H-F: $z=3.48$, $p<10^{-3}$, Cohen's $d=0.47$; H-S vs. H-C: $z=3.72$, $p<10^{-3}$, Cohen's $d=0.73$), there are no significant differences in human cooperation rates when interacting with different types of agents in other intervals.
 Due to the limited number of participants whose normative expectations fall within the $80\%-100\%$ interval, the data of this interval are combined with those of the $60\%-80\%$ interval. Two-tailed Mann–Whitney U tests are used for pairwise comparisons.}  
\label{given norm}    
\end{figure}

\clearpage
\newpage

\begin{figure}[tbh!]
\centering
\includegraphics[width=1\linewidth]{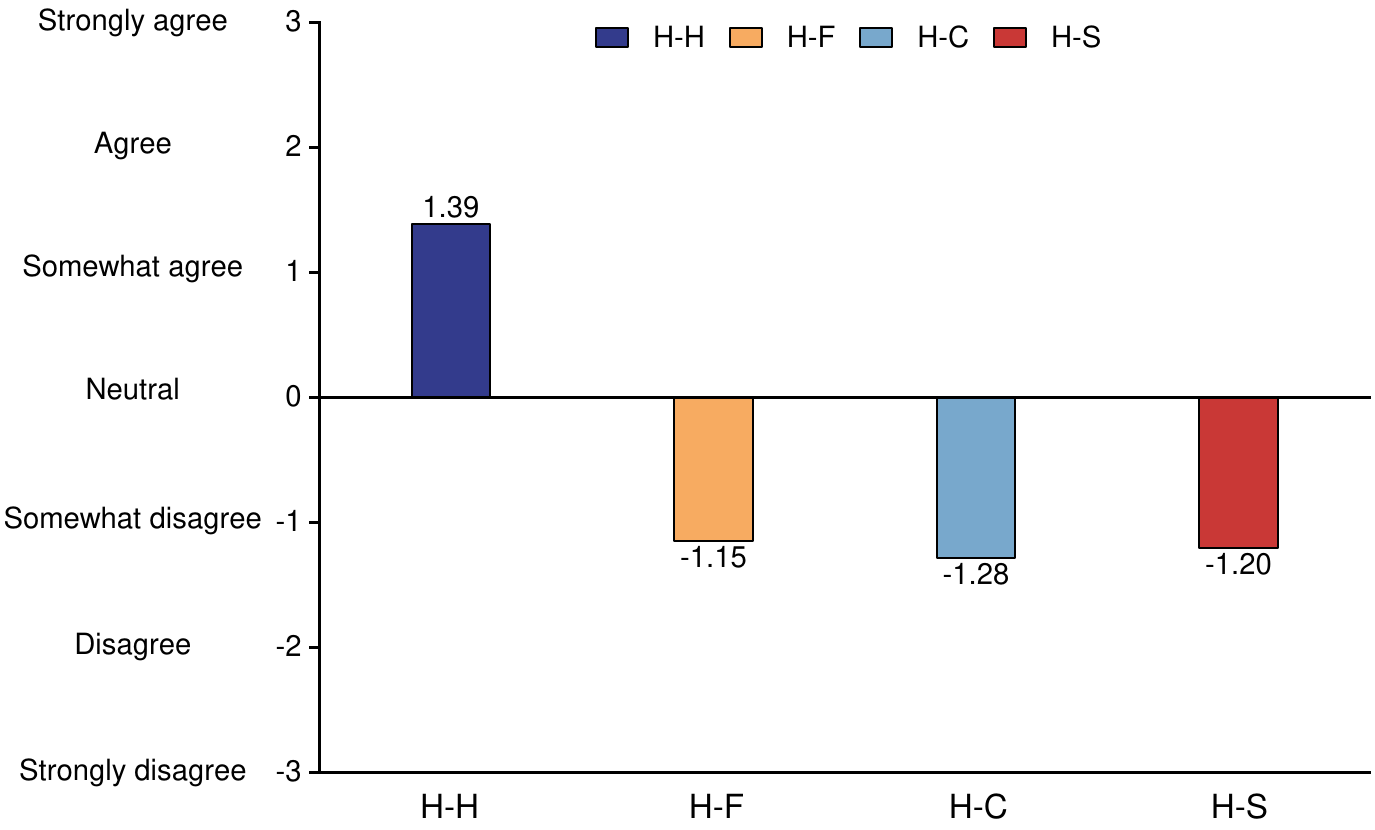}
\caption{\textbf{Humans are able to differentiate between AI agents and humans, when they are aware of the potential involvement of artificial entities under the label-uninformed setting.} 
The panel depicts participants' agreement levels regarding whether their associates are humans or not, collected through a post-experiment questionnaire. In interactions with fellow humans, participants gave significantly higher than zero scores (H-H: $V=7422$, $p<10^{-13}$), indicating that they could accurately tell that their associates were human. In contrast, in interactions with AI agents, participants gave significantly lower than zero scores  (H-F: $V=1489.5$, $p<10^{-10}$; H-C: $V=1642.5$, $p<10^{-10}$; H-S: $V=1812$, $p<10^{-9}$), indicating they could accurately tell that their associates were non-human. The one-sample Wilcoxon signed-rank test is employed to determine whether the mean scores significantly differ from zero.
}  
\label{Turing test}    
\end{figure}

\clearpage
\newpage

\begin{table}[!htbp]
\renewcommand\arraystretch{1.1}
\setlength{\tabcolsep}{16pt}
\centering
\caption{Results of the one-way Analysis of Variance (ANOVA) assessing the impact of treatment types, i.e. human-human, human-cooperative AI agents, human-fair AI agents, and human-selfish AI agents, under the label-informed setting, on human cooperation rates. For detailed post-hoc analysis, please refer to Table \ref{HSD label-informed}.}
\begin{tabular}{llllll}
\hline
Term               & d.f.$^{a}$ & S.S$^{b}$   & MS$^{c}$   & F statistic & $p$-value      \\ \hline
Condition          & 3    & 6.42  & 2.14 & 22.96       & 4.75e-14 *** \\
Residuals          & 572  & 53.27 & 0.09 &             &              \\ \hline
$^{a}$Degrees of freedom &      &       &      &             &              \\
$^{b}$Sum of squares     &      &       &      &             &              \\
$^{c}$Mean squares       &      &       &      &             &              \\ \hline
\end{tabular}
\label{anova label-informed}
\end{table}

\clearpage
\newpage

\begin{table}[!htbp]
\renewcommand\arraystretch{1.1}
\setlength{\tabcolsep}{3pt}
\centering
\caption{
Pairwise comparisons of pre-game cooperation agreements and in-game promise breaches in four treatments under the label-informed setting: human-human (H-H), human-cooperative AI agents (H-C), human-fair AI agents (H-F), and human-selfish AI agents (H-S). 
The table examines the percentages of (1) pre-game cooperation agreements between humans and AI agents, (2) humans’ in-game promise breaches, and (3) AI agents’ in-game promise breaches, respectively.
For each pairwise comparison, the table includes the percentages of the two treatments, the $\chi^{2}$ statistic, and the $p$-value. Note that the subscripts 1 and 2 after the percentage indicate the first and second treatments, respectively.  Statistical test results are obtained through two-sample proportions $Z$ test.}
\begin{tabular}{cccccc}
\hline
Dimension                                        & Treatment   & Percentage1 (\%) & Percentage2 (\%) & $\chi^{2}$     & $p$-value      \\ \hline
\multirow{6}{*}{\begin{tabular}[c]{@{}c@{}}Pre-game\\ Cooperation Agreements\end{tabular}} & H-H vs. H-F & 75.5            & 87.5            & 20.311  & 6.583e-6***  \\
                                                 & H-H vs. H-C & 75.5            & 81.1            & 0.196   & 0.657        \\
                                                 & H-H vs. H-S & 75.5            & 79.7            & 0.009   & 0.924        \\
                                                 & H-F vs. H-C & 87.5            & 81.1            & 22.697  & 1.896e-6***  \\
                                                 & H-F vs. H-S & 87.5            & 79.7            & 30.818  & 2.834e-8***  \\
                                                 & H-C vs. H-S & 81.1            & 79.7            & 0.565   & 0.452        \\ \hline
\multirow{6}{*}{\begin{tabular}[c]{@{}c@{}}Human In-Game \\ Promise Breach\end{tabular}}    & H-H vs. H-F & 45.2            & 56.8            & 31.031  & 2.539e-8***  \\
                                                 & H-H vs. H-C & 45.2            & 68.2            & 120.530 & \textless 2.2e-16*** \\
                                                 & H-H vs. H-S & 45.2            & 79.2            & 273.810 & \textless 2.2e-16*** \\
                                                 & H-F vs. H-C & 56.8            & 68.2            & 32.945  & 9.482e-9***  \\
                                                 & H-F vs. H-S & 56.8            & 79.2            & 135.800 & \textless 2.2e-16*** \\
                                                 & H-C vs. H-S & 68.2            & 79.2            & 35.336  & 2.774e-9***  \\ \hline
\multirow{3}{*}{\begin{tabular}[c]{@{}c@{}}AI Agent In-Game \\ Promise Breach\end{tabular}} & H-F vs. H-C & 9.6             & 0.0             & 115.930 & \textless 2.2e-16*** \\
                                                 & H-F vs. H-S & 9.6             & 71.7            & 968.920 & \textless 2.2e-16*** \\
                                                 & H-C vs. H-S & 0.0             & 71.7            & 1295.0  & \textless 2.2e-16*** \\ \hline
\end{tabular}%
\label{Reach break promise label-informed}
\end{table}

\clearpage
\newpage

\begin{table}[!htbp]
\renewcommand\arraystretch{1.1}
\setlength{\tabcolsep}{5pt}
\centering
\caption{Pairwise comparisons of participants' agreement levels for their associates' personality and mindfulness in four treatments under the label-informed setting: human-human (H-H), human-cooperative AI agents (H-C), human-fair AI agents (H-F), and human-selfish AI agents (H-S). The agreement levels are on 7-point Likert scales, ranging from $-3$ (strong disagreement) to $3$ (strong agreement). For each pairwise comparison, the table includes the median and mean agreement levels of the two treatments, the $W$ statistic, and the $p$-value. Note that the subscripts 1 and 2 after the median or mean indicate the first and second treatments, respectively.  Statistical test results are obtained through two-tailed Mann-Whitney $U$ test.}
\begin{tabular}{llcccccc}
\hline
\textbf{Treatment}          & \textbf{Dimension} & \textbf{Median1} & \textbf{Median2} & \textbf{Mean1} & \textbf{Mean2} & \textbf{W} & \textbf{$p$-value}     \\ \hline
H-H vs. H-F & Trustworthiness    & -1               & 2                & -0.590         & 1.451          & 4378.5     & \textless 2.2e-16*** \\
                            & Intelligence       & 1                & 1                & 0.645          & 0.500          & 11465      & 0.114                \\
                            & Cooperativeness    & -1               & 2                & -0.500         & 1.542          & 4145.5     & \textless 2.2e-16*** \\
                            & Likability         & -1               & 2                & 0.708          & 1.271          & 5047.5     & 2.384e-14***         \\
                            & Fairness           & 0                & 2                & -0.597         & 1.375          & 5721.5     & 2.23e-11***          \\
                            & Agency             & 1                & 1                & 0.250          & 0.653          & 9173       & 0.086                \\
                            & Experience         & 1                & 0                & 0.931          & 0.139          & 13373      & 1.612e-05***         \\ \hline
H-H vs. H-C & Trustworthiness    & -1               & 3                & -0.590         & 2.424          & 1943.5     & \textless 2.2e-16*** \\
                            & Intelligence       & 1                & 0                & 0.646          & -0.104         & 12890      & 2.968e-4***          \\
                            & Cooperativeness    & -1               & 3                & -0.500         & 2.659          & 1349.5     & \textless 2.2e-16***     \\
                            & Likability         & -1               & 2.5              & 0.708          & 2.083          & 2852.5     & \textless 2.2e-16***     \\
                            & Fairness           & 0                & 2                & -0.597         & 1.958          & 3919       & \textless 2.2e-16***     \\
                            & Agency             & 1                & 2                & 0.250          & 1.347          & 6862       & 4.592e-07***         \\
                            & Experience         & 1                & 0                & 0.931          & -0.125         & 13597      & 3.762e-06***         \\ \hline
H-H vs. H-S & Trustworthiness    & -1               & -2               & -0.590         & -1.424         & 12892      & 2.668e-4***          \\
                            & Intelligence       & 1                & 1                & 0.646          & 0.611          & 10890      & 0.451                \\
                            & Cooperativeness    & -1               & -2               & -0.500         & -1.361         & 12712      & 7.08e-4***           \\
                            & Likability         & -1               & -1               & 0.708          & -1.319         & 12859      & 3.352***             \\
                            & Fairness           & 0                & -1               & -0.597         & -1.014         & 13386      & 1.491e-05***         \\
                            & Agency             & 1                & 0                & 0.250          & -0.250         & 11980      & 0.021*               \\
                            & Experience         & 1                & 0                & 0.931          & 0.083          & 13230      & 3.986e-05***         \\ \hline
H-F vs. H-C & Trustworthiness    & 2                & 3                & 1.451          & 2.424          & 6142       & 1.597e-10***         \\
                            & Intelligence       & 1                & 0                & 0.500          & -0.104         & 12226      & 0.008**              \\
                            & Cooperativeness    & 2                & 3                & 1.542          & 2.659          & 5367       & 6.417e-15***         \\
                            & Likability         & 2                & 2.5              & 1.271          & 2.083          & 7109.5     & 1.533e-06***         \\
                            & Fairness           & 2                & 2                & 1.375          & 1.958          & 7754.5     & 1.216***             \\
                            & Agency             & 1                & 2                & 0.653          & 1.347          & 7713       & 1.283***             \\
                            & Experience         & 0                & 0                & 0.139          & -0.125         & 11254      & 0.205                \\ \hline
\end{tabular}
\label{perception label-informed}
\end{table}

\clearpage
\newpage

\begin{table}[!htbp]
\renewcommand\arraystretch{1.1}
\setlength{\tabcolsep}{1.5pt}
\centering
\caption{\textbf{Basic information on the conducted experimental sessions.} In total, 16 sessions were divided between eight treatments. Sessions were characterized by the order of experiment without communication (WoC) and with communication (WC), the number of interactions, attendance, the mean age of participants and its standard deviation, and the percentage of women.
H-H, H-F, H-S, and  H-C represent treatments conducted under the label-uninformed setting.
H-HP, H-FP, H-SP, and  H-CP represent treatments conducted under the label-informed setting.}
\begin{tabular}{ccccccccc}
\hline
Date                            & Treatment            & Location                          & Order  & Interactions & Participants & Mean age & SD age & \%women \\ \hline
10 March 2024          & H-H                 & Xi'an                             & WoC-WC & 10-10        & 72           & 18.9     & 0.69   & 29.1    \\
                                 & H-H                 & Xi'an                             & WC-WoC & 10-10        & 72           & 18.7     & 0.77   & 37.5    \\
9 March 2024            & H-F                 & Xi'an                             & WoC-WC & 10-10        & 72           & 18.7     & 0.81   & 54.1    \\
                                 & H-F                 & Xi'an                             & WC-WoC & 10-10        & 72           & 18.8     & 0.68   & 61.1    \\
20 March 2024           & H-C                 & Taiyuan                           & WoC-WC & 10-10        & 72           & 19.9     & 1.06   & 61.1    \\
                                 & H-C                 & Taiyuan                           & WC-WoC & 10-10        & 72           & 19.7     & 0.85   & 59.7    \\
17 April 2024           & H-S                 & Taiyuan                           & WoC-WC & 10-10        & 72           & 19.1     & 1.04   & 34.7    \\
                                 & H-S                 & Taiyuan                           & WC-WoC & 10-10        & 72           & 19.1     & 1.90   & 37.5    \\
18 May 2024            & H-HP                & Kunming                           & WoC-WC & 10-10        & 72           & 21.9     & 2.49   & 48.6    \\
                                 & H-HP                & Kunming                           & WC-WoC & 10-10        & 72           & 21.1     & 1.99   & 47.2    \\
14,15 March 2024        & H-FP                & Xi'an                             & WoC-WC & 10-10        & 72           & 25.1     & 1.92   & 77.7    \\
                                 & H-FP                & Xi'an                             & WC-WoC & 10-10        & 72           & 19.5     & 1.46   & 58.3    \\
27 April 2024           & H-CP                & Kunming                           & WoC-WC & 10-10        & 72           & 20.9     & 2.03   & 59.7    \\
                                 & H-CP                & Kunming                           & WC-WoC & 10-10        & 72           & 21.1     & 2.28   & 63.8    \\
28 April 2024           & H-SP                & Kunming                           & WoC-WC & 10-10        & 72           & 22.3     & 2.53   & 45.8    \\
                                 & H-SP                & Kunming                           & WC-WoC & 10-10        & 72           & 20.2     & 1.81   & 47.2    \\ \hline
\end{tabular}
\label{basic info}
\end{table}

\clearpage

\begin{table}[!htbp]
\renewcommand\arraystretch{1.1}
\setlength{\tabcolsep}{3pt}
\centering
\caption{Pairwise comparisons of participants' agreement levels for their associates' personality and mindfulness in four treatments under the label-uninformed setting: human-human (H-H), human-cooperative AI agents (H-C), human-fair AI agents (H-F), and human-selfish AI agents (H-S). The agreement levels are on 7-point Likert scales, ranging from $-3$ (strong disagreement) to $3$ (strong agreement). For each pairwise comparison, the table includes the median and mean agreement levels of the two treatments, the $W$ statistic, and the $p$-value. Note that the subscripts 1 and 2 after the median or mean indicate the first and second treatments, respectively.  Statistical test results are obtained through two-tailed Mann-Whitney $U$ test.}
\begin{tabular}{llcccccc}
\hline
\textbf{Treatment}          & \textbf{Dimension} & \textbf{Median1} & \textbf{Median2} & \textbf{Mean1} & \textbf{Mean2} & \textbf{W} & \textbf{$p$-value}     \\ \hline
H-H vs. H-F & Trustworthiness    & -1               & 2                & -0.972         & 1.250          & 3747       & \textless 2.2e-16*** \\
                            & Intelligence       & 1                & 1                & 0.340          & 0.409          & 10144      & 0.748                \\
                            & Cooperativeness    & -1               & 2                & -0.924         & 1.465          & 3447       & \textless 2.2e-16*** \\
                            & Likability         & -1               & 2                & -0.854         & 1.257          & 3932       & \textless 2.2e-16*** \\
                            & Fairness           & 0                & 2                & -0.444         & 1.424          & 4188.5     & \textless 2.2e-16*** \\
                            & Agency             & -1               & 1                & -0.375         & 0.701          & 6908.5     & 6.788e-07***         \\
                            & Experience         & 1                & 0                & 0.208          & 0.194          & 10594      & 0.746                \\ \hline
H-H vs. H-C & Trustworthiness    & -1               & 3                & -0.972         & 2.340          & 1469.5     & \textless 2.2e-16*** \\
                            & Intelligence       & 1                & 1                & 0.340          & 0.465          & 9895       & 0.498                \\
                            & Cooperativeness    & -1               & 3                & -0.924         & 2.542          & 932.5      & \textless 2.2e-16*** \\
                            & Likability         & -1               & 2                & -0.854         & 2.076          & 1847       & \textless 2.2e-16*** \\
                            & Fairness           & 0                & 2                & -0.444         & 2.194          & 1948       & \textless 2.2e-16*** \\
                            & Agency             & -1               & 2                & -0.375         & 1.201          & 5497       & 2.953e-12***         \\
                            & Experience         & 1                & 0.5              & 0.208          & 0.319          & 9952       & 0.552                \\ \hline
H-H vs. H-S & Trustworthiness    & -1               & -2               & -0.972         & -1.743         & 13029      & 1.027e-4***          \\
                            & Intelligence       & 1                & -2               & 0.340          & 0.028          & 11200      & 0.232                \\
                            & Cooperativeness    & -1               & -2               & -0.924         & -1.618         & 12970      & 1.544e-04***         \\
                            & Likability         & -1               & -2               & -0.854         & -1.340         & 12224      & 0.007**              \\
                            & Fairness           & 0                & -1               & -0.444         & -0.549         & 10884      & 0.459                \\
                            & Agency             & -1               & -1               & -0.375         & -0.563         & 11002      & 0.363                \\
                            & Experience         & 1                & 0                & 0.208          & -0.167         & 11630      & 0.070                \\ \hline
H-F vs. H-C & Trustworthiness    & 2                & 3                & 1.250          & 2.340          & 5538.5     & 7.116e-13***         \\
                            & Intelligence       & 1                & 1                & 0.409          & 0.465          & 10130      & 0.733                \\
                            & Cooperativeness    & 2                & 3                & 1.465          & 2.542          & 5988.5     & 2.926e-11***         \\
                            & Likability         & 2                & 2                & 1.257          & 2.076          & 7302       & 6.308e-06***         \\
                            & Fairness           & 2                & 2                & 1.424          & 2.194          & 7186       & 2.327e-06***         \\
                            & Agency             & 1                & 2                & 0.701          & 1.201          & 8479       & 0.006**              \\
                            & Experience         & 0                & 0.5              & 0.194          & 0.319          & 9912       & 0.514                \\ \hline
\end{tabular}
\label{perception label-uninformed}
\end{table}

\clearpage
\newpage
\begin{table}[!htbp]
\renewcommand\arraystretch{1.1}
\setlength{\tabcolsep}{5pt}
\centering
\caption{Pairwise comparisons of participants' agreement levels for their associates' communication quality in four treatments under the label-uninformed setting: 
human-human (H-H), human-cooperative AI agents (H-C), human-fair AI agents (H-F), and human-selfish AI agents (H-S). The agreement levels are on 7-point Likert scales, ranging from $-3$ (strong disagreement) to $3$ (strong agreement). For each pairwise comparison, the table includes the median and mean agreement levels of the two treatments, the $W$ statistic, and the $p$-value. Note that the subscripts 1 and 2 after the median or mean indicate the first and second treatments, respectively.  Statistical test results are obtained through two-tailed Mann-Whitney $U$ test.}
\begin{tabular}{llcccccc}
\hline
\textbf{Treatment}          & \textbf{Dimension} & \textbf{Median1} & \textbf{Median2} & \textbf{Mean1} & \textbf{Mean2} & \textbf{W} & \textbf{$p$-value}     \\ \hline
H-H vs. H-F & Clarity            & 1                & 2                & 0.396          & 2.097          & 4774.5     & 4.199e-16***         \\
                            & Conciseness        & 1                & 1                & 0.965          & 0.229          & 12780      & 5.045e-04***         \\
                            & Concreteness       & 1                & 2                & 0.056          & 1.840          & 4290       & \textless 2.2e-16*** \\
                            & Coherence          & 1                & 2                & 0.159          & 1.507          & 5617       & 6.588e-12***         \\
                            & Courteousness      & 1                & 2                & 0.889          & 2.208          & 5318       & 9.035e-14***         \\
                            & Correctness        & 1.5              & 2                & 1.104          & 1.715          & 7897.5     & 3.067e-04***         \\
                            & Completeness       & 1                & 2                & 0.986          & 1.840          & 7206.5     & 3.664e-06***         \\ \hline
H-H vs. H-C & Clarity            & 1                & 3                & 0.396          & 2.409          & 3491.5     & \textless 2.2e-16*** \\
                            & Conciseness        & 1                & 1                & 0.965          & 0.688          & 11136      & 0.266                \\
                            & Concreteness       & 1                & 2                & 0.056          & 2.007          & 3570       & \textless 2.2e-16*** \\
                            & Coherence          & 1                & 2                & 0.159          & 1.715          & 4836.5     & 1.343e-15***         \\
                            & Courteousness      & 1                & 2.5              & 0.889          & 2.361          & 4564.5     & \textless 2.2e-16*** \\
                            & Correctness        & 1.5              & 2                & 1.104          & 1.819          & 7321.5     & 8.254e-06***         \\
                            & Completeness       & 1                & 2                & 0.986          & 1.993          & 6430       & 8.061e-09***         \\ \hline
H-H vs. H-S & Clarity            & 1                & 2                & 0.396          & 1.354          & 7117.5     & 2.722e-06***         \\
                            & Conciseness        & 1                & 1                & 0.965          & 0.264          & 12293      & 0.006**              \\
                            & Concreteness       & 1                & 1                & 0.056          & 0.972          & 7125.5     & 2.867e-06***         \\
                            & Coherence          & 1                & 1                & 0.159          & 0.417          & 9293.5     & 0.121                \\
                            & Courteousness      & 1                & 2                & 0.889          & 1.792          & 6953.5     & 5.682e-07***         \\
                            & Correctness        & 1.5              & 2                & 1.104          & 1.479          & 8819.5     & 0.024*               \\
                            & Completeness       & 1                & 2                & 0.986          & 1.646          & 7885       & 2.910e-04***         \\ \hline
H-F vs. H-C & Clarity            & 2                & 3                & 2.097          & 2.409          & 8388.5     & 0.002**              \\
                            & Conciseness        & 1                & 1                & 0.229          & 0.688          & 8816.5     & 0.025*               \\
                            & Concreteness       & 2                & 2                & 1.840          & 2.007          & 9229       & 0.089                \\
                            & Coherence          & 2                & 2                & 1.507          & 1.715          & 9293.5     & 0.114                \\
                            & Courteousness      & 2                & 2.5              & 2.208          & 2.361          & 9204.5     & 0.072                \\
                            & Correctness        & 2                & 2                & 1.715          & 1.819          & 9739       & 0.351                \\
                            & Completeness       & 2                & 2                & 1.840          & 1.993          & 9325.5     & 0.119                \\ \hline
\end{tabular}
\label{SI Table communication}
\end{table}

\clearpage
\newpage

\begin{table}[!htbp]
\renewcommand\arraystretch{1.1}
\setlength{\tabcolsep}{2pt}
\centering
\caption{Generalized linear models under the label-uninformed setting that take participants' cooperation rates as dependent variables, and various aspects of perceptions of AI agents, collected through post-experiment questionnaires, as independent variables. Separate models are constructed for the human-human treatment and the human-AI (H-C, H-S, and H-F) treatments, with the H-F treatment serving as the baseline. The generalized linear model indicates that normative expectation is the most influential factor, regardless of whether participants interact with human or AI agent associates. However, the impact of other factors on human cooperation differs between human-human and human-AI treatments. In human-human treatment, the perceived likability, cooperativeness, as well as communication conciseness, clarity, courteousness, and completeness, are significant predictors. In contrast, for the human-AI treatments, the perceived intelligence, experience, cooperativeness, likability, agency of AI agents, along with communication completeness, concreteness, and correctness, are significant predictors.}
\begin{tabular}{ccccclcccc}
\hline
\multicolumn{1}{l}{}   & \multicolumn{9}{c}{Dependent Variable: Human Cooperation Rates}                        \\ \hline
Model & \multicolumn{4}{c}{Humans vs. Humans}     &  & \multicolumn{4}{c}{Humans vs. AI agents}         \\  
 &
  \textit{Coef.$^a$} &
  \textit{S.E.$^b$} &
  \textit{z} value &
  pr(\textgreater{} $\lvert z \rvert$) &
   &
  \textit{Coef.$^a$} &
  \textit{S.E.$^b$} &
  \textit{z value} &
  pr(\textgreater{} $\lvert z \rvert$) \\ \hline
Intercept              & -0.60                & 0.06                 & -10.01                & \textless 2e-16 ***  &  & -1.09 & 0.07 & -15.21 & \textless 2e-16 *** \\
Clarity                & -0.49                 & 0.09                 & -5.39                 & 6.75e-8 ***           &  & 0.08 & 0.05 & 1.73  & 0.08                \\
Conciseness            & 0.25                & 0.08                 & 3.26                & 1.13e-3 **                 &  & 0.04 & 0.04 & 1.02  & 0.31              \\
Concreteness           & 0.18                 & 0.09                 & 1.79                 & 0.07                 &  & -0.10 & 0.05 & -1.99  & 0.04 *                \\
Coherence              & 0.09                & 0.09                 & 1.09                & 0.28                 &  & 0.09 & 0.06 & 1.75  & 0.08               \\
Courteousness          & -0.26                & 0.08                 & -3.19                & 1.42e-3 **                 &  & -0.05 & 0.05 & -0.97  & 0.33                \\
Correctness            & 0.16                & 0.08                 & 1.90                & 0.06                 &  & -0.18  & 0.06 & -3.28   & 1.05e-03 **                \\
Completeness           & -0.27                & 0.08                 & -3.39                & 6.95e-4 ***                 &  & 0.14 & 0.06 & 2.38  & 0.02 *                \\
Trustworthiness        & -0.07                 & 0.11                 & -0.66                 & 0.51           &  & 0.19 & 0.09 & 1.87  & 0.06                \\
Intelligence           & 0.19                & 0.08                 & 2.49                & 0.01 *                 &  & 0.39  & 0.05 & 7.99   & 1.36e-15 ***        \\
Cooperativeness        & 0.39                 & 0.12                 & 3.39                 & 6.79e-4 ***            &  & -0.39 & 0.11 & -3.75  & 1.76e-4 ***                \\
Likability             & -0.24                & 0.12                 & -2.01                & 0.04 *                 &  & -0.24  & 0.09 & -2.79   & 5.31e-3 **               \\
Fairness               & -0.19                 & 0.09                 & -1.95                 & 0.05                 &  & 0.05  & 0.07 & 0.68   & 0.49          \\
Agency                 & -0.11                & 0.08                 & -1.37                & 0.17                 &  & -0.11 & 0.06 & -2.04  & 0.04 *                \\
Experience             & 0.08                & 0.08                 & 1.01                & 0.31                 &  & 0.15  & 0.05 & 3.01   & 2.57e-3 **               \\
Normative Expectation  & 0.66                 & 0.07                 & 9.24                & \textless 2e-16 ***  &  & 1.06  & 0.05 & 21.93  & \textless 2e-16 *** \\
Treatment Effect H-C   & \multicolumn{1}{l}{} & \multicolumn{1}{l}{} & \multicolumn{1}{l}{} & \multicolumn{1}{l}{} &  & -0.02  & 0.09 & -0.17   & 0.87                \\
Treatment Effect H-S   & \multicolumn{1}{l}{} & \multicolumn{1}{l}{} & \multicolumn{1}{l}{} & \multicolumn{1}{l}{} &  & 0.11 & 0.13 & 0.82  & 0.41               \\
Null deviance          & \multicolumn{4}{c}{650.9}                                                                 &  & \multicolumn{4}{c}{2569.9}                  \\
Residual deviance      & \multicolumn{4}{c}{434.2}                                                                 &  & \multicolumn{4}{c}{1679.0}                  \\
AIC$^c$ & \multicolumn{4}{c}{751.5}                                                                 &  & \multicolumn{4}{c}{2348.4}                  \\
Obeservation           & \multicolumn{4}{c}{144}                                                                   &  & \multicolumn{4}{c}{432}                     \\ \hline
\multicolumn{1}{l}{$^a$Coefficient} &
  \multicolumn{1}{l}{} &
  \multicolumn{1}{l}{} &
  \multicolumn{1}{l}{} &
  \multicolumn{1}{l}{} &
   &
  \multicolumn{1}{l}{} &
  \multicolumn{1}{l}{} &
  \multicolumn{1}{l}{} &
  \multicolumn{1}{l}{} \\
\multicolumn{1}{l}{$^b$Standard error} &
  \multicolumn{1}{l}{} &
  \multicolumn{1}{l}{} &
  \multicolumn{1}{l}{} &
  \multicolumn{1}{l}{} &
   &
  \multicolumn{1}{l}{} &
  \multicolumn{1}{l}{} &
  \multicolumn{1}{l}{} &
  \multicolumn{1}{l}{} \\
\multicolumn{1}{l}{$^c$Akaike information criterion} &
  \multicolumn{1}{l}{} &
  \multicolumn{1}{l}{} &
  \multicolumn{1}{l}{} &
  \multicolumn{1}{l}{} &
   &
  \multicolumn{1}{l}{} &
  \multicolumn{1}{l}{} &
  \multicolumn{1}{l}{} &
  \multicolumn{1}{l}{} \\ \hline
\end{tabular}

\label{GLM Questionnaire}
\end{table}

\clearpage
\newpage

\begin{table}[!htbp]
\renewcommand\arraystretch{1.1}
\setlength{\tabcolsep}{5pt}
\centering
\caption{Pairwise comparisons of participants' agreement levels for their associates' communication quality in four treatments under the label-informed setting: 
human-human (H-H), human-cooperative AI agents (H-C), human-fair AI agents (H-F), and human-selfish AI agents (H-S). The agreement levels are on 7-point Likert scales, ranging from $-3$ (strong disagreement) to $3$ (strong agreement). For each pairwise comparison, the table includes the median and mean agreement levels of the two treatments, the $W$ statistic, and the $p$-value. Note that the subscripts 1 and 2 after the median or mean indicate the first and second treatments, respectively.  Statistical test results are obtained through two-tailed Mann-Whitney $U$ test.}
\begin{tabular}{llcccccc}
\hline
\textbf{Treatment}          & \textbf{Dimension} & \textbf{Median1} & \textbf{Median2} & \textbf{Mean1} & \textbf{Mean2} & \textbf{W} & \textbf{$p$-value} \\ \hline
H-H vs. H-F & Clarity            & 2                & 2                & 1.319          & 1.764          & 8738       & 0.017*           \\
                            & Conciseness        & 2                & 1                & 1.354          & 1.159          & 11468      & 0.106            \\
                            & Concreteness       & 1                & 2                & 0.951          & 1.868          & 7328.5     & 8.04e-06***      \\
                            & Coherence          & 2                & 2                & 1.111          & 1.451          & 9462.5     & 0.185            \\
                            & Courteousness      & 2                & 2                & 1.326          & 2.000          & 7939       & 3.175e-4***      \\
                            & Correctness        & 2                & 2                & 1.639          & 1.493          & 11125      & 0.265            \\
                            & Completeness       & 2                & 2                & 1.465          & 1.799          & 9214.5     & 0.086            \\ \hline
H-H vs. H-C & Clarity            & 2                & 3                & 1.319          & 2.486          & 5574.5     & 6.443e-13***     \\
                            & Conciseness        & 2                & 2                & 1.354          & 1.618          & 9256       & 0.102            \\
                            & Concreteness       & 1                & 2                & 0.951          & 2.201          & 5681.5     & 4.594e-12***     \\
                            & Coherence          & 2                & 2                & 1.111          & 1.750          & 7917.5     & 3.324e-4***      \\
                            & Courteousness      & 2                & 3                & 1.326          & 2.535          & 5050       & 1.297e-15***     \\
                            & Correctness        & 2                & 2                & 1.639          & 1.972          & 8636       & 0.009**          \\
                            & Completeness       & 2                & 2                & 1.465          & 2.201          & 6941       & 2.845e-07***     \\ \hline
H-H vs. H-S & Clarity            & 2                & 2                & 1.319          & 1.083          & 11013      & 0.349            \\
                            & Conciseness        & 2                & 1                & 1.354          & 0.986          & 11862      & 0.029*           \\
                            & Concreteness       & 1                & 2                & 0.951          & 1.236          & 9645       & 0.292            \\
                            & Coherence          & 2                & 1                & 1.111          & 0.750          & 11924      & 0.024*           \\
                            & Courteousness      & 2                & 2                & 1.326          & 1.472          & 10202      & 0.809            \\
                            & Correctness        & 2                & 2                & 1.639          & 1.500          & 11251      & 0.191            \\
                            & Completeness       & 2                & 2                & 1.465          & 1.556          & 10491      & 0.855            \\ \hline
H-F vs. H-C & Clarity            & 2                & 3                & 1.764          & 2.486          & 7076.5     & 4.955e-07***     \\
                            & Conciseness        & 1                & 2                & 1.159          & 1.618          & 8157       & 0.001**          \\
                            & Concreteness       & 2                & 2                & 1.868          & 2.201          & 8324.5     & 0.002**          \\
                            & Coherence          & 2                & 2                & 1.451          & 1.750          & 8614.5     & 0.009**          \\
                            & Courteousness      & 2                & 3                & 2.000          & 2.535          & 6914       & 9.869e-08***     \\
                            & Correctness        & 2                & 2                & 1.493          & 1.972          & 7900.5     & 2.557e-04***     \\
                            & Completeness       & 2                & 2                & 1.799          & 2.201          & 7905.5     & 1.948e-04***     \\ \hline
\end{tabular}
\label{communication label-informed}
\end{table}

\clearpage
\newpage

\begin{table}[!htbp]
\renewcommand\arraystretch{1.1}
\setlength{\tabcolsep}{16pt}
\centering
\caption{Results of the one-way Analysis of Variance (ANOVA) assessing the impact of treatment types i.e. human-human, human-cooperative AI agents, human-fair AI agents, and human-selfish AI agents, under the label-uninformed setting, on human cooperation rates. For detailed post-hoc analysis, please refer to SI, Table \ref{s1 table1b}.}
\begin{tabular}{llllll}
\hline
Term               & d.f.$^{a}$ & S.S$^{b}$   & MS$^{c}$   & F statistic & $p$-value      \\ \hline
Condition          & 3    & 1.30  & 0.4341 & 4.277       & 0.005 ** \\
Residuals          & 572  & 58.06 & 0.1015 &             &              \\ \hline
$^{a}$Degrees of freedom &      &       &      &             &              \\
$^{b}$Sum of squares     &      &       &      &             &              \\
$^{c}$Mean squares       &      &       &      &             &              \\ \hline
\end{tabular}
\label{s1 table1}
\end{table}

\clearpage
\newpage

\begin{table}[htb]
\renewcommand\arraystretch{1.1}
\setlength{\tabcolsep}{18pt}
\centering
\caption{Summary of Tukey Honest Significant Difference (HSD) post-hoc analysis for the impact of treatment types i.e. human-human (H-H), human-cooperative AI agents (H-C), human-fair AI agents (H-F), and human-selfish AI agents (H-S), under the label-uninformed setting, on human cooperation rates. The table shows for each pairwise comparison, the difference in the mean human cooperation rate, the lower confidence bound (LCB), the upper confidence bound (UCB), and the $p$-value. There is no significant difference between H-F and H-H treatments, H-F and H-C treatments, H-H and H-C treatments, H-S and H-C treatments. However, there are significant differences between  H-S and H-F treatments, as well as H-S and H-H treatments. }
\begin{tabular}{ccccccc}
\hline
\multicolumn{3}{c}{Comparison} & Difference & LCB$^a$    & UCB$^b$    & $p$-value            \\ \hline
H-F      & vs.      & H-C      & 0.041      & -0.056  & 0.138  & 0.695            \\
H-H      & vs.      & H-C      & 0.063      & -0.034  & 0.159  & 0.334      \\
H-S      & vs.      & H-C      & -0.062     & -0.159 & 0.035 & 0.354             \\
H-H      & vs.      & H-F      & 0.022      & -0.075 & 0.119  & 0.935              \\
H-S      & vs.      & H-F      & -0.103     & -0.199 & -0.006 & 0.032 *     \\
H-S      & vs.      & H-H      & -0.125     & -0.222 & -0.029 & 0.005 ** \\ \hline
\multicolumn{7}{l}{$^a$Lower confidence bound}                                         \\
\multicolumn{7}{l}{$^b$Upper confidence bound}                                         \\ \hline
\end{tabular}
\label{s1 table1b}
\end{table}

\clearpage
\newpage

\begin{table}[!htbp]
\renewcommand\arraystretch{1.1}
\setlength{\tabcolsep}{3pt}
\centering
\caption{Pairwise comparisons of pre-game cooperation agreements and in-game promise breaches in four treatments under the label-uninformed setting: human-human (H-H), human-cooperative AI agents (H-C), human-fair AI agents (H-F), and human-selfish AI agents (H-S). 
The table examines the percentages of (1) pre-game cooperation agreements between humans and AI agents, (2) humans’ in-game promise breaches, and (3) AI agents’ in-game promise breaches, respectively.
For each pairwise comparison, the table includes the percentages of the two treatments, the $\chi^{2}$ statistic, and the $p$-value. Note that the subscripts 1 and 2 after the percentage indicate the first and second treatments, respectively.  Statistical test results are obtained through two-sample proportions $Z$ test.}
\begin{tabular}{cccccc}
\hline
Dimension                                        & Treatment   & Percentage1 (\%) & Percentage2 (\%) & $\chi^{2}$     & $p$-value      \\ \hline
\multirow{6}{*}{\begin{tabular}[c]{@{}c@{}}Pre-game\\ Cooperation Agreements\end{tabular}} & H-H vs. H-F & 66.9            & 92.2            & 138.380  & \textless 2.2e-16***  \\
                                                 & H-H vs. H-C & 66.9            & 86.7            & 57.101   & 4.139e-14***        \\
                                                 & H-H vs. H-S & 66.9            & 73.9            & 0.024   & 0.876        \\
                                                 & H-F vs. H-C & 92.2            & 86.7            & 22.440  & 2.168e-6***  \\
                                                 & H-F vs. H-S & 92.2            & 73.9            & 158.560  & \textless 2.2e-16***  \\
                                                 & H-C vs. H-S & 86.7            & 73.9            & 74.348   & \textless 2.2e-16***        \\ \hline
\multirow{6}{*}{\begin{tabular}[c]{@{}c@{}}Human In-Game \\ Promise Breach\end{tabular}}    & H-H vs. H-F & 54.5            & 63.0            & 16.886  & 3.969e-5***  \\
                                                 & H-H vs. H-C & 54.5            & 66.5            & 33.147 & 8.546e-9*** \\
                                                 & H-H vs. H-S & 54.5            & 70.9            & 58.835 & 1.715e-14*** \\
                                                 & H-F vs. H-C & 63.0            & 66.5            & 3.210  & 0.073  \\
                                                 & H-F vs. H-S & 63.0            & 70.9            & 16.047 & 6.178e-5*** \\
                                                 & H-C vs. H-S & 66.5            & 70.9            & 4.978  & 0.026*  \\ \hline
\multirow{3}{*}{\begin{tabular}[c]{@{}c@{}}AI Agent In-Game \\ Promise Breach\end{tabular}} & H-F vs. H-C & 12.6             & 0.2             & 160.010 & \textless 2.2e-16*** \\
                                                 & H-F vs. H-S & 12.6             & 72.4            & 883.120 & \textless 2.2e-16*** \\
                                                 & H-C vs. H-S & 0.2             & 72.4            & 1344.0  & \textless 2.2e-16*** \\ \hline
\end{tabular}%
\label{Reach break promise label-uninformed}
\end{table}

\clearpage
\newpage

\begin{table}[htb]
\renewcommand\arraystretch{1.1}
\setlength{\tabcolsep}{17pt}
\centering
\caption{Summary of Tukey Honest Significant Difference (HSD) post-hoc analysis for the impact of treatment types, i.e. human-human (H-H), human-cooperative AI agents (H-C), human-fair AI agents (H-F), and human-selfish AI agents (H-S), under the label-informed setting, on human cooperation rates. The table shows for each pairwise comparison, the difference in the mean human cooperation rate, the lower confidence bound (LCB), the upper confidence bound (UCB), and the $p$-value. There is no significant difference between H-F and H-H treatments. However, there are significant differences between H-F and H-C treatments, H-H and H-C treatments, H-S and H-C treatments, H-S and H-F treatments, as well as H-S and H-H treatments. }
\begin{tabular}{ccccccc}
\hline
\multicolumn{3}{c}{Comparison} & Difference & LCB$^a$    & UCB$^b$    & $p$-value            \\ \hline
H-F      & vs.      & H-C      & 0.131      & 0.038  & 0.223  & 0.002 **           \\
H-H      & vs.      & H-C      & 0.169      & 0.076  & 0.261  & 2.010e-05 ***      \\
H-S      & vs.      & H-C      & -0.097     & -0.189 & -0.004 & 0.037*             \\
H-H      & vs.      & H-F      & 0.038      & -0.054 & 0.131  & 0.713              \\
H-S      & vs.      & H-F      & -0.227     & -0.319 & -0.134 & \textless 0.001 ***     \\
H-S      & vs.      & H-H      & -0.265     & -0.358 & -0.173 & \textless 0.001 *** \\ \hline
\multicolumn{7}{l}{$^a$Lower confidence bound}                                         \\
\multicolumn{7}{l}{$^b$Upper confidence bound}                                         \\ \hline
\end{tabular}
\label{HSD label-informed}
\end{table}

\clearpage
\newpage

\begin{table}[!t]
\renewcommand\arraystretch{1.1}
\setlength{\tabcolsep}{3pt}
\centering
\caption{Generalized linear models under the label-informed setting that take participants' cooperation rates as dependent variables,  and various aspects of perceptions of AI agents, collected through post-experiment questionnaires, as independent variables. Separate models are constructed for the human-human treatment and the human-AI (H-C, H-S, and H-F) treatments, with the H-F treatment serving as the baseline. The generalized linear model indicates that normative expectation is the most influential factor, regardless of whether participants interact with human or agent associates. However, the impact of other factors on human cooperation varies between human-human and human-AI treatments. In human-human treatment, the perceived trustworthiness and clarity of communication from fellow humans are significant predictors. In contrast, for the human-AI treatments, the perceived intelligence and fairness of AI agents, along with message conciseness are significant predictors. }
\begin{tabular}{ccccclcccc}
\hline
\multicolumn{1}{l}{}   & \multicolumn{9}{c}{Dependent Variable: Human Cooperation Rates}                        \\ \hline
Model & \multicolumn{4}{c}{Humans vs. Humans}     &  & \multicolumn{4}{c}{Humans vs. AI agents}         \\   
 &
  \textit{Coef.$^a$} &
  \textit{S.E.$^b$} &
  \textit{z} value &
  pr(\textgreater{} $\lvert z \rvert$) &
   &
  \textit{Coef.$^a$} &
  \textit{S.E.$^b$} &
  \textit{z value} &
  pr(\textgreater{} $\lvert z \rvert$) \\ \hline
Intercept              & -0.19                & 0.06                 & -3.32                & 9.17e-4 ***          &  & -0.95 & 0.07 & -13.64 & \textless 2e-16 *** \\
Clarity                & 0.24                 & 0.08                 & 3.19                 & 1.42e-3 **           &  & -0.03 & 0.05 & -0.56  & 0.58                \\
Conciseness            & -0.03                & 0.07                 & -0.40                & 0.69                 &  & -0.10 & 0.04 & -2.23  & 0.03 *              \\
Concreteness           & 0.04                 & 0.08                 & 0.45                 & 0.65                 &  & -0.02 & 0.06 & -0.39  & 0.70                \\
Coherence              & -0.14                & 0.09                 & -1.59                & 0.11                 &  & -0.05 & 0.05 & -0.93  & 0.35                \\
Courteousness          & -0.11                & 0.08                 & -1.38                & 0.17                 &  & -0.09 & 0.05 & -1.82  & 0.07                \\
Correctness            & -0.04                & 0.08                 & -0.48                & 0.63                 &  & 0.03  & 0.05 & 0.66   & 0.51                \\
Completeness           & -0.01                & 0.07                 & -0.19                & 0.85                 &  & -0.10 & 0.05 & -1.75  & 0.08                \\
Trustworthiness        & 0.37                 & 0.11                 & 3.19                 & 1.43e-3 **           &  & -0.04 & 0.11 & -0.40  & 0.69                \\
Intelligence           & -0.13                & 0.07                 & -1.82                & 0.07                 &  & 0.39  & 0.05 & 8.14   & 3.95e-16 ***        \\
Cooperativeness        & 0.01                 & 0.14                 & 0.06                 & 0.95                 &  & -0.12 & 0.12 & -0.99  & 0.32                \\
Likability             & -0.22                & 0.13                 & -1.70                & 0.09                 &  & 0.01  & 0.10 & 0.11   & 0.91                \\
Fairness               & 0.03                 & 0.09                 & 0.34                 & 0.73                 &  & 0.23  & 0.08 & 3.00   & 2.74e-3 **          \\
Agency                 & -0.05                & 0.08                 & -0.58                & 0.56                 &  & -0.08 & 0.05 & -1.53  & 0.13                \\
Experience             & -0.01                & 0.07                 & -0.10                & 0.92                 &  & 0.07  & 0.05 & 1.54   & 0.12                \\
Normative Expectation  & 0.72                 & 0.07                 & 10.37                & \textless 2e-16 ***  &  & 0.85  & 0.04 & 19.03  & \textless 2e-16 *** \\
Treatment Effect H-C   & \multicolumn{1}{l}{} & \multicolumn{1}{l}{} & \multicolumn{1}{l}{} & \multicolumn{1}{l}{} &  & 0.04  & 0.10 & 0.43   & 0.67                \\
Treatment Effect H-S   & \multicolumn{1}{l}{} & \multicolumn{1}{l}{} & \multicolumn{1}{l}{} & \multicolumn{1}{l}{} &  & -0.31 & 0.13 & -2.36  & 0.02 *              \\
Null deviance          & \multicolumn{4}{c}{786.8}                                                                 &  & \multicolumn{4}{c}{2284.1}                  \\
Residual deviance      & \multicolumn{4}{c}{565.8}                                                                 &  & \multicolumn{4}{c}{1388.5}                  \\
AIC$^c$ & \multicolumn{4}{c}{869.9}                                                                 &  & \multicolumn{4}{c}{2120.4}                  \\
Obeservation           & \multicolumn{4}{c}{144}                                                                   &  & \multicolumn{4}{c}{432}                     \\ \hline
\multicolumn{1}{l}{$^a$Coefficient} &
  \multicolumn{1}{l}{} &
  \multicolumn{1}{l}{} &
  \multicolumn{1}{l}{} &
  \multicolumn{1}{l}{} &
   &
  \multicolumn{1}{l}{} &
  \multicolumn{1}{l}{} &
  \multicolumn{1}{l}{} &
  \multicolumn{1}{l}{} \\
\multicolumn{1}{l}{$^b$Standard error} &
  \multicolumn{1}{l}{} &
  \multicolumn{1}{l}{} &
  \multicolumn{1}{l}{} &
  \multicolumn{1}{l}{} &
   &
  \multicolumn{1}{l}{} &
  \multicolumn{1}{l}{} &
  \multicolumn{1}{l}{} &
  \multicolumn{1}{l}{} \\
\multicolumn{1}{l}{$^c$Akaike information criterion} &
  \multicolumn{1}{l}{} &
  \multicolumn{1}{l}{} &
  \multicolumn{1}{l}{} &
  \multicolumn{1}{l}{} &
   &
  \multicolumn{1}{l}{} &
  \multicolumn{1}{l}{} &
  \multicolumn{1}{l}{} &
  \multicolumn{1}{l}{} \\ \hline
\end{tabular}
\label{GLM Questionnaire label-informed}
\end{table}

\clearpage
\newpage



\end{document}